\documentclass[fleqn,twocolumn]{aastex631}
\DeclareRobustCommand{\VAN}[3]{#2}
\let\VANthebibliography\thebibliography
\def\thebibliography{\DeclareRobustCommand{\VAN}[3]{##3}\VANthebibliography}

\newcommand{\oneE}{1E~1547.0$-$5408}    
\newcommand{\LS}{LS~5039}    
\newcommand{\NuS}{NuSTAR}
\newcommand{\ASCA}{ASCA}
\newcommand{\Su}{Suzaku}
\newcommand{\ax}{a_{\rm x}\sin i}
\newcommand{\zz}{Z_4^2}
\newcommand{\Eb}{E_{\rm b}}
\newcommand{\R}{R_{\rm ppd}}
\newcommand{\Pns}{P_{\rm NS}}
\newcommand{\Pch}{{\cal P}_{\rm ch}}
\newcommand{\chisq}{\chi_{\rm r}}

\usepackage{graphicx}	
\usepackage{amsmath}	
\usepackage{amssymb}	
\usepackage{booktabs}
\begin{document}
\received{2023/8/5}
\accepted{2023/11/10}
\title{Further Evidence for the $\sim 9$ s Pulsation in LS 5039 from NuSTAR and ASCA}
\author{Kazuo  Makishima}
\affiliation{Department of Physics, The University of Tokyo,
7-3-1 Hongo, Bunkyo-ku, Tokyo, 113-0033, Japan }
\affiliation{
Kavli Institute for the Physics and Mathematics of the Universe (WPI),\\
The University of Tokyo,
5-1-5 Kashiwa-no-ha, Kashiwa, Chiba, 277-8683, Japan}
\author{Nagomi Uchida}
\affiliation{Institute of Space and Astronautical Science,
3-1-1 Yoshinodai, Chuo-ku, Sagamihara, Kanagawa, 252-5210, Japan}
\author{Hiroki Yoneda}
\affiliation{Department of Physics and Astronomy, University of W\"urzburg,
Sanderring 2, 97070 W\"urzburg,Germany }
\author{Teruaki Enoto}
\affiliation{Department of Physics, Kyoto University,
Kitashirakawa Oiwake-cho, Sakyo-ku, Kyoto, 606-8502, Japan }
\affiliation{Extreme Natural Phenomena RIKEN Hakubi Research Team,
Cluster for Pioneering Research, RIKEN, \\
2-1 Hirosawa, Wako, Saitama, 351-0198, Japan }
\author{Tadayuki Takahashi}
\affiliation{
Kavli Institute for the Physics and Mathematics of the Universe (WPI),\\
The University of Tokyo,
5-1-5 Kashiwa-no-ha, Kashiwa, Chiba 277-8683, Japan}
%
\begin{abstract}
The  present  study aims to reinforce the evidence 
for the $\sim 9$ s pulsation in the gamma-ray binary \LS,
derived  with a \Su\ observation in 2007 and that with \NuS\  in 2016
\citep{Yoneda20}.
Through a reanalysis of the \NuS\ data
incorporating the orbital Doppler correction,
the 9.0538 s pulsation was confirmed successfully even  in the 3--10 keV range,
where it was undetectable  previously.
This was attained by  perceiving an energy-dependent 
drift in the pulse phase below 10 keV,
and correcting the pulse timing of individual photons for that effect.
Similarly, an archival 0.7--12 keV data set of \LS,
taken with the \ASCA\ GIS in 1999 October, was analyzed.
The data showed possible periodicity at about 8.882 s,
but again the energy-dependent phase drift was noticed below 10 keV.
By correcting for this effect,
and for the orbital Doppler delays in the \LS\ system,
the 2.8--12 keV periodicity became statistically significant 
at $8.891 \pm 0.001$ s.
The periods measured with \ASCA, \Su, and \NuS\ 
approximately follow an average period derivative 
of $\dot P \approx 3.0 \times 10^{-10}$ s s$^{-1}$.
These results provide further evidence for the pulsation in this object,
and strengthen the scenario by \cite{Yoneda20},
that the compact object in \LS\  is a strongly magnetized neutron star.
\end{abstract}
\email{maxima@phys.s.u-tokyo.ac.jp}
\keywords{Astrophysical magnetism --- Gamma-ray sources
---Magnetars  --- Neutron Stars --- Pulsars}

\section{INTRODUCTION}
\label{sec:intro}
Through hard X-ray observations with \Su\ in 2007 and \NuS\ in 2016, \cite{Yoneda20},
hereafter Paper I, derived evidence for  a $\sim 9$ s pulsation  from \LS,
the prototypical gamma-ray binary.
This apparently settles the controversy \citep[e.g.,][]{Dubus13}
about the nature of the compact object  in this system,
indicating that it is a neutron star (NS), rather than a black hole.
The result further led \cite{Yoneda20,Yoneda21} to propose
that this NS is in fact has a magnetar-class strong magnetic fields,
which are responsible for the particle acceleration
and the consequent bright MeV gamma-ray emission from \LS.
It is also suggested that magnetars can reside,
not only as isolated objects, but also in binaries.

In spite of these rich implications,
Paper I left several basic issues unsolved.
(i) The pulsation in the \NuS\ data
may not have solid significance yet.
In fact, \citet{Volkov21} reconfirmed the \Su\ pulses
at the same period as in Paper I (see Table~\ref{tbl:orbital_parameters}),
but failed to confirm the pulsation in the present \NuS\ data.
(ii) For some unknown reasons,
the pulses become  invisible in the \NuS\ data below 10 keV,
even though no particular feature is seen in the X-ray continuum at $\sim 10$ keV
\citep{Takahashi09,Kishishita09,Volkov21}.
(The \Su\ HXD data are limited to $>10$ keV.)
This problem is indeed puzzling, 
and is invoked as an argument against the reality of 
the $\sim 9$ s pulsation \citep{Volkov21}.
(iii) The orbital solutions derived with \Su\ and \NuS\ are not yet 
fully consistent with each other,
particularly in the eccentricity $e$ and the X-ray semi-major axis $\ax$.
(iv) The 10--30 keV pulse fraction is much higher in the HXD data,
$0.68 \pm 0.14$, than in the \NuS\ data,
$0.135 \pm 0.043$.

In the present paper, we aim at solving  (i) and (ii) above.
They are to some extent coupled;
any hint of the same pulsation,
if confirmed in softer energies, 
will enhance the pulse reliability,
and open a way to the pulse searches
utilizing archival soft X-ray data of \LS.
As to (ii), understanding the origin of the pulse disappearance below 10 keV 
would provide important information as to the X-ray emission mechanism of this object.

For the above purposes,
we first reanalyze,  in \S~\ref{sec:NuSTAR_reanalysis},
the same \NuS\ data as used in Paper I and \cite{Volkov21},
but  focusing on energies below 10 keV.
Based on that result,
we analyze in \S~\ref{sec:GIS_analysis}
the 0.7--12 keV data of \LS\ 
acquired in 1999 with the \ASCA\ GIS, 
hoping to reconfirm the  pulsation.
In these studies, searches for periodic signals are 
carried out utilizing the standard epoch-folding method.
The epoch-folded profiles are 
examined for the periodicity significance,
employing the $Z_m^2$ statistics \citep{Z2_94, Makishima16}
with the harmonic number set to either $m=2$ or $m=4$ (Paper I).
Since $Z_m^2$ is less widely known 
than the more standard chi-square technique,
in Appendix A we compare the two statistics,
and show that $Z_m^2$  much more suited to the present aims.
The orbital period of  \LS\ is fixed at 3.90608 d 
\citep{Aragona09}.
All errors refer to 68\% confidence limits.

\section{REANALYSIS OF THE NuSTAR DATA}
\label{sec:NuSTAR_reanalysis}

\begin{figure}
\centerline{
\includegraphics[width=8.5cm]{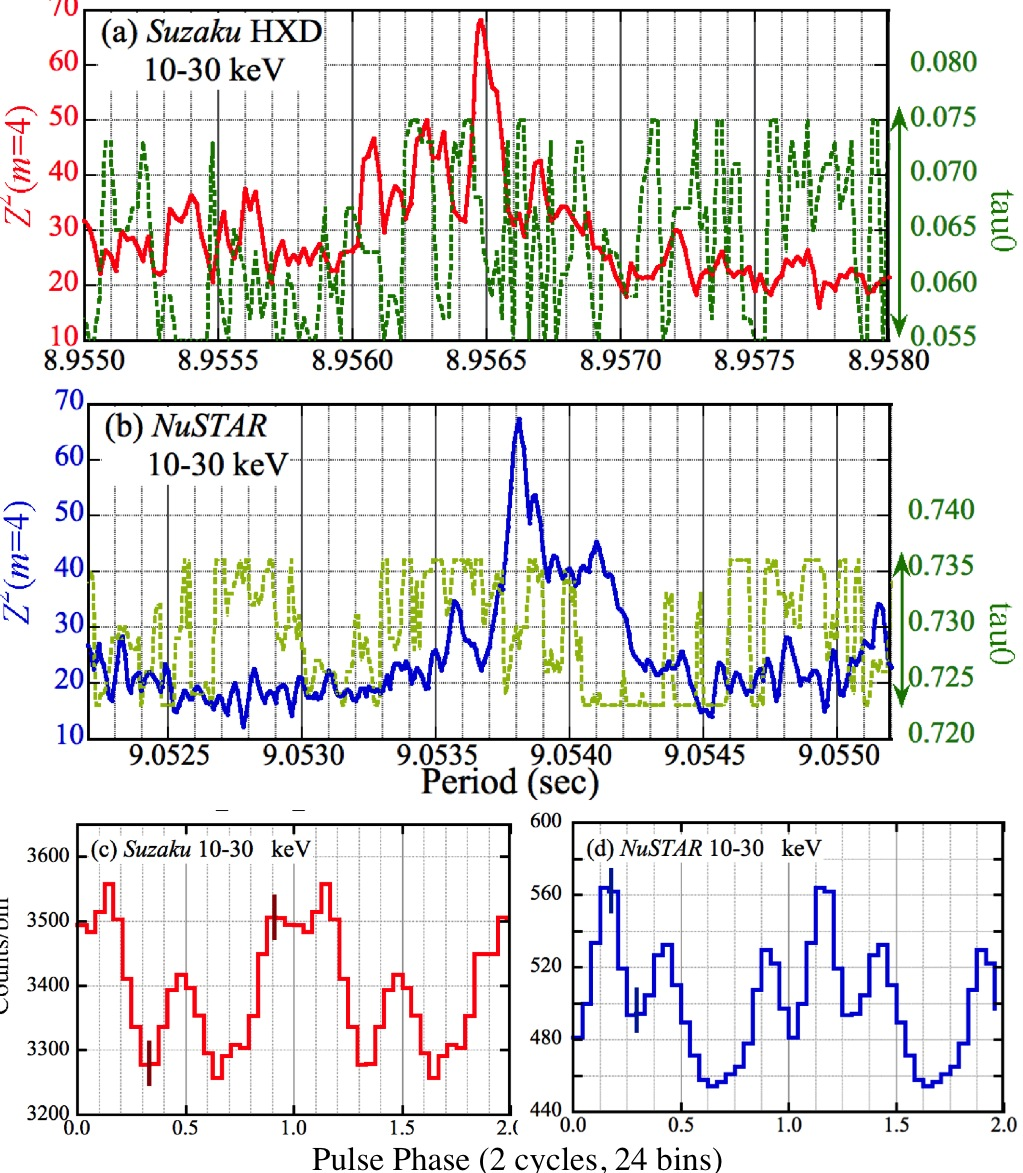}
}
\vspace*{-2mm}
\caption{Essentials of Paper I. 
(a) and (b) are PGs from the 10--30 keV
 \Su\ and \NuS\ data, respectively.
 They are corrected for the orbit of \LS,
with the orbital parameters  readjusted at each period (see text).
Red/blue shows $\zz$ with $m=4$ (left ordinate), 
whereas the dashed green curve gives $\tau_0$ (right ordinate),
with its search range indicated by a  green arrow.
(c) and (d) are background-inclusive pulse profiles,
folded using the orbital solutions
specified by the $\zz$ peak
in (a) and (b), respectively. 
Between  (c) and (d), the relative pulse phase is arbitrary. 
}
\label{fig:Paper_I_review}
\end{figure}

\subsection{A brief review of Paper I}
\label{subsec_PaperI_review}

To begin with, let us briefly review the  results  in Paper I.
Although  the observation and basic data reduction described therein are not repeated here,
some important numbers  are reproduced in Table~\ref{tbl:orbital_parameters}. 
There, we present not only the $\zz$ values, 
but also the reduced chi-square values, $\chisq$,
calculated  with 20 bins ($\nu=19$ degrees of freedom) and 16 bins ($\nu=15$).

The $\zz$ periodograms (PGs) in the 10--30 keV energy range, 
from   \Su\ and \NuS,
are reproduced in Figures~\ref{fig:Paper_I_review}(a) 
and  \ref{fig:Paper_I_review}(b), respectively.
Like in Figure~3 of Paper I,
they are Doppler corrected,
using the orbital solutions obtained in Paper I.
However, unlike in Paper I,
at each period we readjusted
$\ax$,  the argument of perigee $\omega$, 
and the initial orbital phase $\tau_0$ 
(zero when the X-ray object is at the periastron),
instead of fixing them to the optimum single values.
These parameters were varied over their respect error ranges  (Paper I),
while $e$ is fixed at 0.306 for (a) and 0.278 for (b).
The PG peaks represent the pulse periods on these occasions
(Table~\ref{tbl:orbital_parameters}).
For the  \NuS\ observation, it is given as
\begin{equation}
\Pns = 9.05381(3)~~{\rm s}.
\label{eq:P_NuSTAR}
\end{equation}
Due to the parameter readjustment,
the peaks in Figure~\ref{fig:Paper_I_review}
are  broader than in Figure~3 of Paper I,
and their widths  faithfully represent 
the uncertainties in $P$ (Paper I).
We regard Figure~\ref{fig:Paper_I_review}(b) 
as a reference when analyzing the \NuS\ data below 10 keV.

Panels (c) and (d) of Figure~\ref{fig:Paper_I_review} reproduce 
the 10--30 keV pulse profiles from \Su\ and \NuS, respectively,
folded using their  best orbital solutions.
They are identical to Figure 2 of Paper I, 
except binning and definition of the pulse-phase origin.
Although the two profiles are somewhat dissimilar,  
they both comprise three peaks per cycle.
Here and hereafter, we show pulse profiles 
after applying a  running average,
with weights of 1/4, 1/2, and 1/4 for consecutive three bins.
As a result, the statistical error in each bin becomes
0.61 times the Poissonian  fluctuation
\citep{Makishima21a}.

\begin{table*} 
\begin{center}
\caption{The orbital and PPD parameters of \LS, derived with \Su,  \NuS, and \ASCA.$^{a)}$}
\label{tbl:orbital_parameters}
\begin{tabular}{llccccccccc}
\hline 
\multicolumn{2}{l}{Energy }  &$e$&  $ \ax $ & $ \omega $ &$ \tau_0 ^{b)}$ & $P$   & $E_{\rm b}$  & $\R$ & $Z_4^2$&$ \chi^2_{\rm r}$ \\
&  (keV)               &                     &    (lt-s)      & (deg)       & ( $0 \le \tau_0 \le 1)$                &  (s)  & (keV)   & (deg keV$^{-1}$)& & ($\nu=19/15$) \\
\hline 
\hline 
\multicolumn{6}{l}{\Su\ HXD (MJD  54352.7163)$^{c)}$} \\
&10--30$^{d)}$ & $0.278^{+0.014}_{-0.023}$  &  $53.05 ^{+0.70}_{-0.55}$    &  $54.6^{+5.1}_{-3.3}$   &  $ 0.067 ^{+0.009}_{-0.012}$    &8.95648(4) 
   & --- & ---  & 67.97&  4.15/5.10\\
 \hline 
\multicolumn{5}{l}{\NuS\ (MJD 57632.0952)$^{c)}$}  \\
&10--30$^{d)}$  &$ 0.306^{+0.015}_{-0.013}$  
                                                   &  $48.1 \pm 0.4$  & $56.8^{+2.3}_{-3.1}$ &  $0.7285^{+0.0078}_{-0.0058}$ & 9.05381(3) & 
                                                   ---  & ---  & 66.87 &  4.24/5.06\\
 &   3.0--9.0    &  (0.306)    &   $48.2 \pm 0.6$    &  $56.3^{+2.5}_{-3.7}$   &  $ 0.730^{+0.008}_{-0.004}$    &9.05385(5) &
    (10.0) & $62 \pm 4$    & 36.22  &  2.51/2.48\\
&3.0--30    &  (0.306)    &   $48.2 \pm 0.5$    &  $56.4^{+2.5}_{-3.3}$   &  $ 0.729^{+0.008}_{-0.005}$    &9.05383(4) &
    $10.4 \pm 0.3$ & $64 \pm 2$    &61.51 & 3.29/4.00\\
\hline 
\multicolumn{7}{l}{\ASCA\ GIS (MJD 51455.529)$^{c)}$} \\
&2.8--12$^{e)}$&  (0.306)    &   $48.2 \pm 1.2$    &  $57.5\pm 1.4$   &  $ 0.44 \pm 0.04$    &8.891(1)  &$10.0^{+2.0}_{-2.3}$&$-23.4^{+2.6}_{-3.7}$& 50.7 &  3.66/3.45\\
&2.8--12$^{e)}$ &  (0.278)    &   $52.0 \pm 1.5$    &  $55.8\pm 1.4$   &  $ 0.43 \pm 0.04$    &8.892(1)  &$10.1^{+2.0}_{-2.2}$&$-23.6^{+2.9}_{-3.5}$& 49.9&  3.41/2.91\\
\hline 
\end{tabular}
\end{center}
\begin{itemize}
\setlength{\itemsep}{-0.3mm}
  \item[$^{a)}$]: The parameters  in parentheses are fixed in the analysis. 
   \item[$^{b)}$]: The initial orbital phase $\tau_0$ is specific to each observation, 
   and does not need to coincide among the three observations.
  \item[$^{c)}$]: The Modified Julian Date of the first event in the data, corresponding to $\tau_0.$
  \item[$^{d)}$]: The parameter values of these rows are taken  from Paper I.
  \item[$^{e)}$]: Errors in these rows do not take into account the uncertainties in $e$.
\end{itemize}
\end{table*}

\subsection{A glance at the data below 10 keV}
\label{subsec_NuS_glance}

We proceed to the \NuS\ data analysis
in  soft X-ray energies below 10 keV.
Starting from the fiducial 10--30 keV PG in Figure~\ref{fig:Paper_I_review}(b),
the lower energy boundary $E_{\rm L}$ was gradually decreased,
and the PG calculation was repeated.
Then, the maximum $\zz$  at 
Equation~(\ref{eq:P_NuSTAR}) quickly decreased, 
from $66.9$ in Figure~\ref{fig:Paper_I_review}(b), 
to $51.5$ at $E_{\rm L}=8$ keV,
and $28.7$ at $E_{\rm L}=6$ keV.
Likewise, no major peaks exceeding $\zz \approx 22$ were seen 
in the 7--10 keV or 5--7 keV PGs,
even when expanding the period range to $P=9.05-9.06$ s.
We thus reconfirm  a conclusion of Paper I, 
that the pulsation disappears below 10 keV of the \NuS\ data.
As described there, a typical pulse fraction is $\approx 0.14$ and  $< 0.03$, 
in energies above and below 10 keV, respectively.

\begin{figure}
\centerline{
\includegraphics[width=9cm]{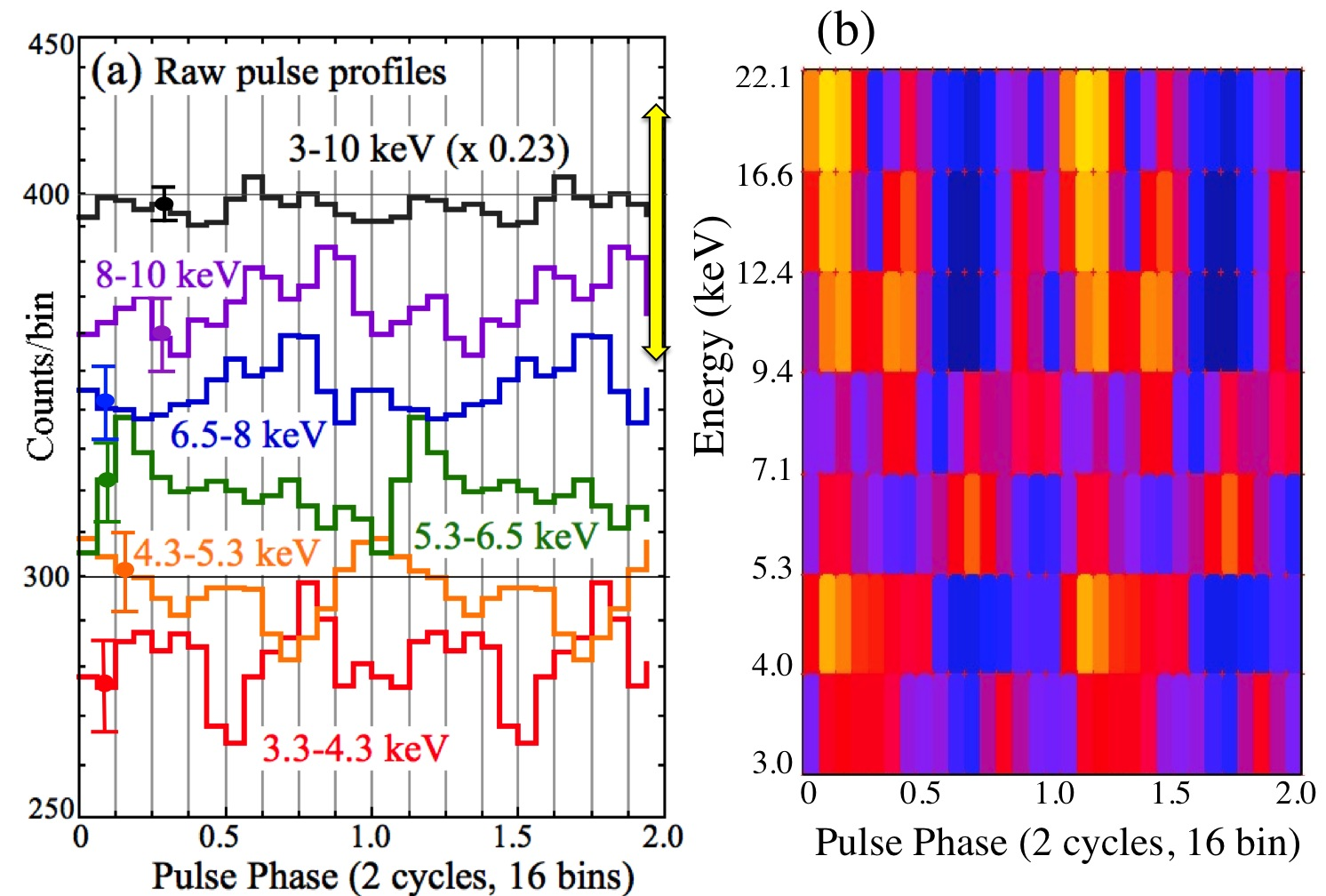}
}
\caption{
(a) Pulse profiles of LS 5039 from the  \NuS\ data,
obtained  in 5 energy bands below 10 keV,
using the same orbital solution as used in Figure~\ref{fig:Paper_I_review}(d).
The ordinate is logarithmic, to make the pulse fraction directly comparable.
A yellow arrow indicates the pulse peak/bottom ratio 
in the 10--30 keV profile (Figure~\ref{fig:Paper_I_review}d).
(b) Photons detected with \NuS\ in 3.0--22.1 keV,
sorted into two dimensions.
Abscissa is  the same  pulse phase  as in (a),
while ordinate is the energy in logarithmic intervals
which increase by a factor of 1.33. 
The color coding is black-blue-purple-red-orange-yellow, from lower to higher counts. 
}
\label{fig:NuSTAR_PulseProfiles_raw}
\end{figure}

To see what is taking place in softer energies,
we next folded the \NuS\ data in several energy bands below 10 keV,
employing the same orbital solution 
as used to produce  Figure~\ref{fig:Paper_I_review}(d).
The results given  in Figure~\ref{fig:NuSTAR_PulseProfiles_raw}(a)
visualize the pulse suppression in lower energies;
the pulse amplitudes, if any,  are 
much lower than that in 10--30 keV (vertical yellow arrow).
Nevertheless, the characteristic ``three-peak" structure 
revealed  in Figure~\ref{fig:Paper_I_review}(d)
still remains, with reduced amplitudes,
in  the 8--10 keV and 6.5--8 keV profiles.
Furthermore, from  6.5--8 keV to 8--10 keV, 
the peaks appear to shift toward later pulse phase. 
Similar ``hard-lag'' behavior is also seen 
between the 3.3--4.3 keV  and  4.3--5.3 keV profiles.

For a further confirmation, 
Figure~\ref{fig:NuSTAR_PulseProfiles_raw}(b) accumulates 
3.0--22.1 keV photons in two dimensions;
the pulse phase (abscissa) and the energy (ordinate), 
where the $n$-th ($n=1, 2, .,7$) energy bin has
$E_{\rm L} = 3.0 \times (1.33)^{n-1}$ keV.
To  make the result  easier to grasp, 
each row of the plot is rescaled to have a mean of 0 
and a standard deviation of 0.5, 
so the pulse-fraction information is lost. 
Above $\sim 10$ keV, we find a clear pulse ridge running vertically, 
accompanied by the two sub peaks at about $\pm 0.25$ cycles off.
These features reconfirm the pulse profile in Figure~\ref{fig:Paper_I_review}(d).
Below 10 keV,  in contrast,
red inclined stripes are seen to run toward lower left,
in agreement with the suggestions of
 Figure~\ref{fig:NuSTAR_PulseProfiles_raw}(a).

Based on these features of Figure~\ref{fig:NuSTAR_PulseProfiles_raw},
we  infer that the pulsation in \LS\ is present in $<10$ keV as well,
but its epoch folding is hampered 
by an energy-dependent  systematic shift  in the pulse phase.
Indeed, from 9 keV to 3 keV, the pulse phase appears 
to tour almost  a complete cycle.
Hereafter, we call this phenomenon  ``Pulse-Phase Drift'',
or  PPD for short, and  regard it as a potential  cause
of the pulse non-detection  below 10 keV.

\subsection{PPD corrections}
\label{subsec:PPD_corr}

\subsubsection{Formalism}

To quantify  the PPD effect,
we  modify the folding analysis;
below an assumed ``break energy" $\Eb $,
the pulse phase $\psi~(0 \le \psi <1)$ is linearly shifted as 
\begin{equation}
\psi'(E)=\psi+ \frac{R _{\rm ppd}}{360}(\Eb - E) ~~~({\rm for}~ E < \Eb) .
\label{eq:PPD}
\end{equation}
where $E$ is the photon energy in keV, 
and $R _{\rm ppd}$ is a coefficient in units of deg keV$^{-1}$.
The pulse phase at $E > \Eb$ is unchanged. 
A positive value of  $R _{\rm ppd}$ means  
corrections for a ``hard lag'', like in the present case,
wherein a lower-energy photon is given a larger  phase delay.
Figure~\ref{fig:NuSTAR_PulseProfiles_raw}(b)
provides an initial guess of $\Eb \sim 10$ keV,
and $R _{\rm ppd}\sim 360^\circ$/(6 keV) =+ 60 deg keV$^{-1}$.

We first attempted to constrain $R _{\rm ppd}$.
To decouple it from $\Eb$,
which is likely to be above $9.4$ keV as in
Figure~\ref{fig:NuSTAR_PulseProfiles_raw}(b),
we chose the 3--9 keV interval and performed the epoch-folding analysis, 
incorporating the correction  by Equation~(\ref{eq:PPD})
in which $E _{\rm b}$ is tentatively fixed  at 10.0 keV.
Scanning  $\R$ between 
$-90$  and $+90$ (1.5 times wider than the above estimate)
with a step of  $1.0$ (all in units of deg keV$^{-1}$), 
we studied how the PG peak at $P=\Pns$   evolves.
In each step of $\R$, we scanned
$\tau_0$ from 0.723 to 0.736 (step 0.001), 
$\ax$ from 47.5 to 48.5 lt-s (step 0.2), 
and $\omega$ from $53^\circ.5$  to $59^\circ.1$  (step $0^\circ.2$),
but $e=0.306$ is fixed.
These ranges are the same as employed in 
calculating Figure~\ref{fig:Paper_I_review}(b).
The pulse period was varied over $\Pns \pm 60~\mu$s,
with a step of $20~\mu$s.

\begin{figure}
\centerline{
\includegraphics[width=8.0cm]{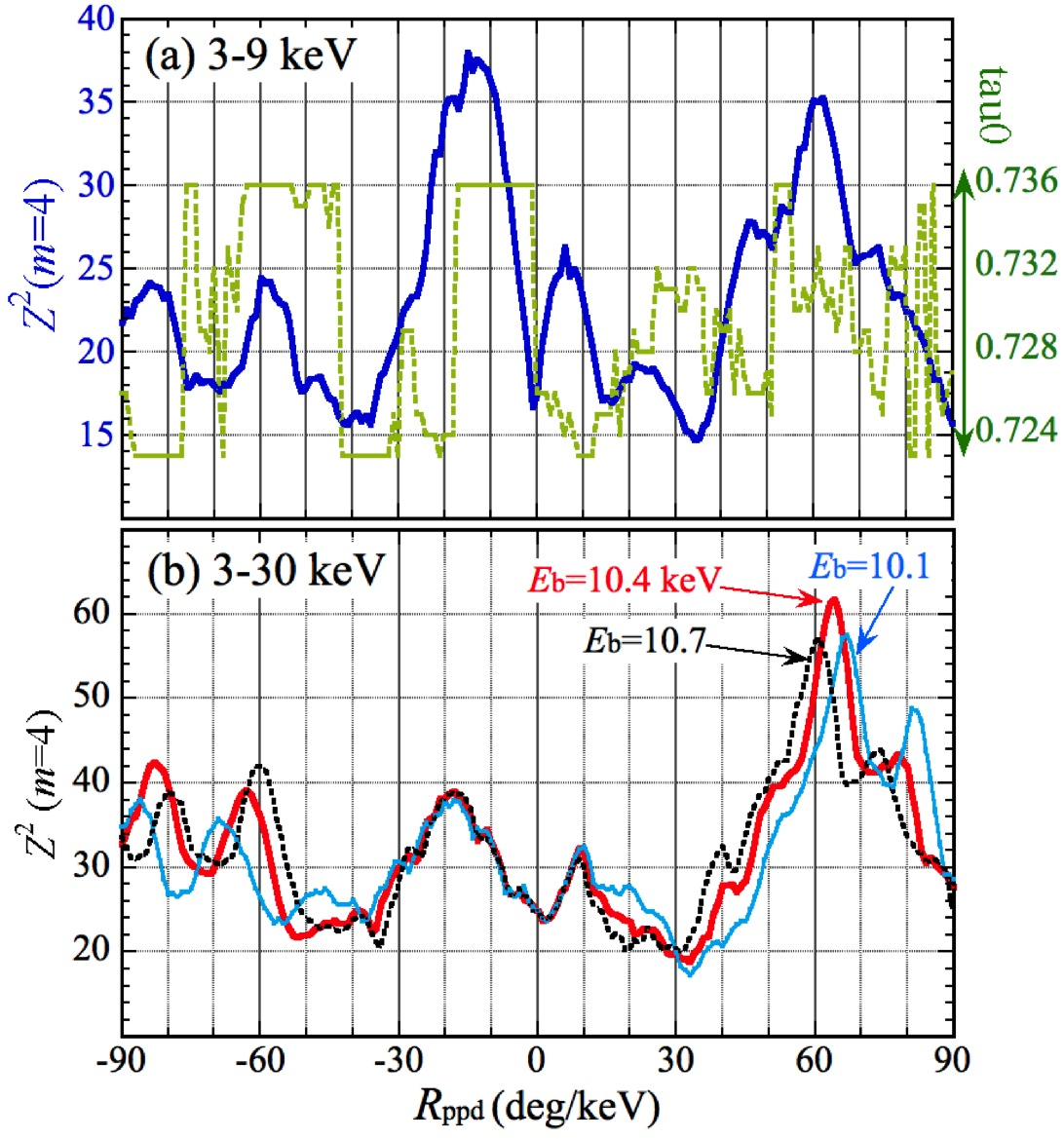}
}
\caption{\NuS\  pulse significance in $Z_4^2$,
shown against  $\R$ in Equation~(\ref{eq:PPD}).
See text for details.
(a) Results in  3--9 keV,
where $Z_4^2$ is in blue  and  $\tau_0$ is in green.
(b) The same, but in 3--30 keV,
shown for three values of $\Eb$;
10.4 keV (solid red), 10.7 keV (dotted black),
and 10.1 keV (cyan).}
\label{fig:NuSTAR_Rscan}
\end{figure}

The result of this analysis is presented in  Figure~\ref{fig:NuSTAR_Rscan}(a).
Strong enhancements in $\zz$ emerged at  
$R _{\rm ppd} \approx 62$  and $\approx -15$ deg keV$^{-1}$,
with the latter somewhat higher.
Compared to $\zz=17.51$ at  $\R=0$ (i.e., no PPD correction),
the peaks at positive and negative $\R$ are higher 
by $\Delta \zz=17.6$ and 20.2, respectively,
where $\Delta \zz$ means increment in $\zz$;
in the relevant parameter range,
$\Delta \zz  \approx+ 11$ means a decrease 
in the chance occurrence probability by two orders of magnitude.
Thus, the correction  by Equation~(\ref{eq:PPD})
appears to be working indeed,
yielding  two candidates of $R _{\rm ppd}$.

\begin{figure}
\centerline{
\includegraphics[width=9.5cm]{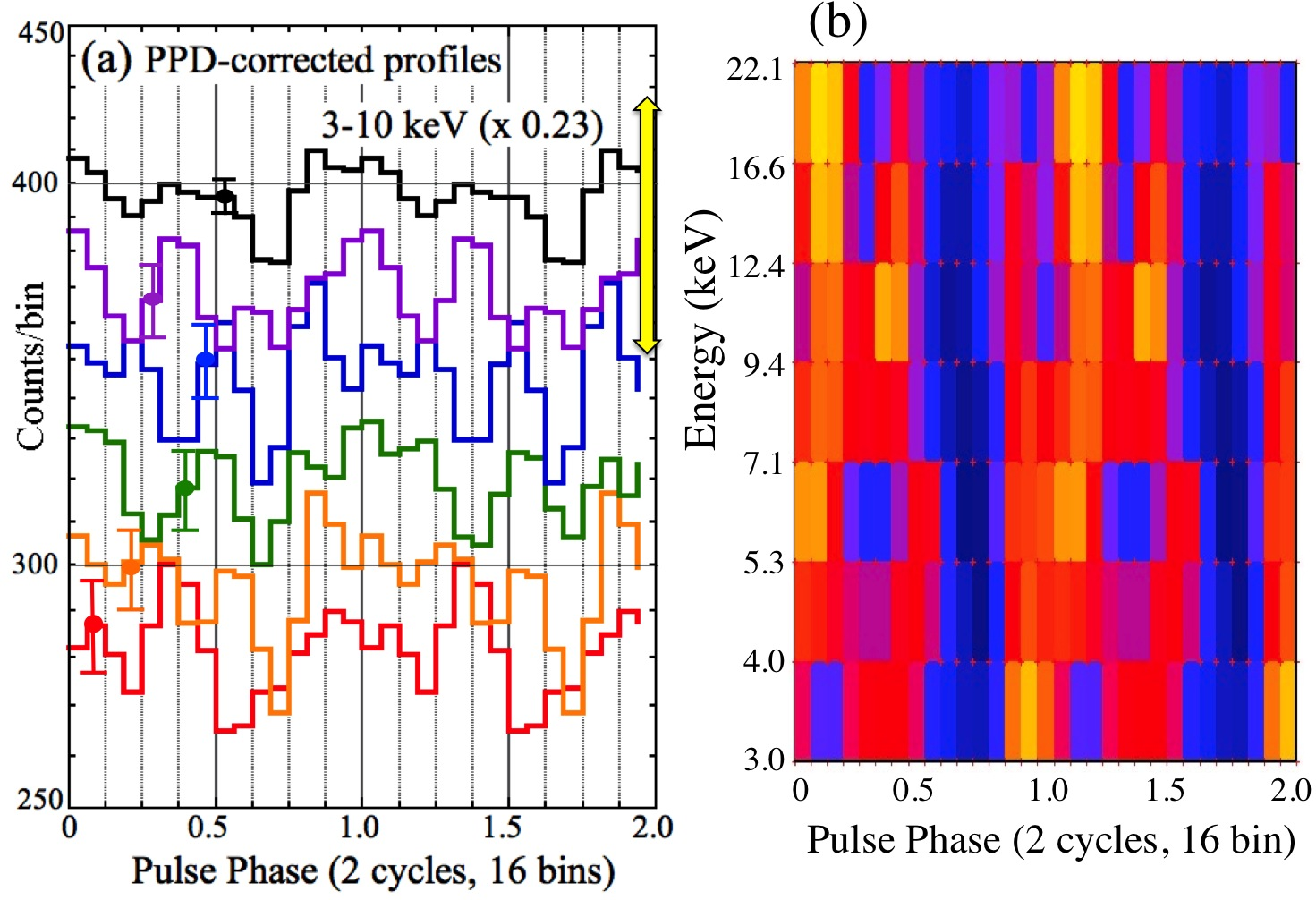}
}
\caption{
The same as Figure~\ref{fig:NuSTAR_PulseProfiles_raw},
but the energy-dependent pulse-phase corrections
with Equation~(\ref{eq:PPD}) have been conducted, 
employing $\Eb=10.4$ keV and $R=63$ deg keV$^{-1}$.}
\label{fig:NuSTAR_PulseProfiles_PPD}
\end{figure}

To estimate  $\Eb$, and decide between 
the two $R _{\rm ppd}$ candidates,
the same analysis was repeated with the 
energy range expanded to 3--30 keV.
By testing several values of  $\Eb$ from 9 keV to 11 keV,
we obtained Figure~\ref{fig:NuSTAR_Rscan}(b).
The positive-$\R$ peak, which is now at  
$\R= 64\pm 2$ deg keV$^{-1}$, increased markedly,
{whereas  the other candidate diminished.
The mechanism working here is  instructive.
The PPD corrections with $\R \approx -15$ and  $\approx 62$
both successfully rectified the $<10$ keV pulse phases.
Compared to the 10--30 keV pulse template,
the soft X-ray  profile derived with $\R=64$ is 
in a relatively good phase alignment,
but that using $\R=-15$
failed to meet this condition. 
As a result, the positive-$\R$ candidate grew up
whereas the other  became weaker,
when we include the 10--30 keV photons
which themselves  should be insensitive to $\R$.
Further considering that 
Figure~\ref{fig:NuSTAR_PulseProfiles_raw}(b) suggests $\R >0$,
and that  the value of  $\tau_0$ (in green) associated with the
positive-$\R$ peak  in Figure~\ref{fig:NuSTAR_Rscan}(a) 
is closer to the Paper I result, 
we adopt  this positive-$R _{\rm ppd}$ solution.
By trimming $\Eb$
and repeating the calculation as 
in Figure~\ref{fig:NuSTAR_Rscan}(b),
we obtained an estimate of  $\Eb = 10.4 \pm 0.3$ keV.

\subsubsection{Results of the PPD correction}
\label{sibsec:NuS_PPD_results}
When the PPD correction  as determined above 
is applied to the photon arrival phases, 
and the orbital parameters are readjusted 
only slightly (Table~\ref{tbl:orbital_parameters}),
Figure~\ref{fig:NuSTAR_PulseProfiles_raw} changed into 
Figure~\ref{fig:NuSTAR_PulseProfiles_PPD}.}
In panel (a), the pulse profiles have become  richer in fine structures,
and they no longer drift with energy.
Likewise, in panel (b), the pulse ridges run
mostly straight throughout the broad energy band.
The 3--10 keV pulse fraction increased to $0.042 \pm 0.022$,
though still much lower than in 10--30 keV
(Figure~\ref{fig:NuSTAR_PulseProfiles_raw}d),
and the profiles remain somewhat different  from the 10-30 keV one.
These are either  intrinsic to the source,
or due to an inaccuracy of Equation~(\ref{eq:PPD})
in modeling the PPD effect.

\begin{figure}
\centerline{
\includegraphics[width=8.6cm]{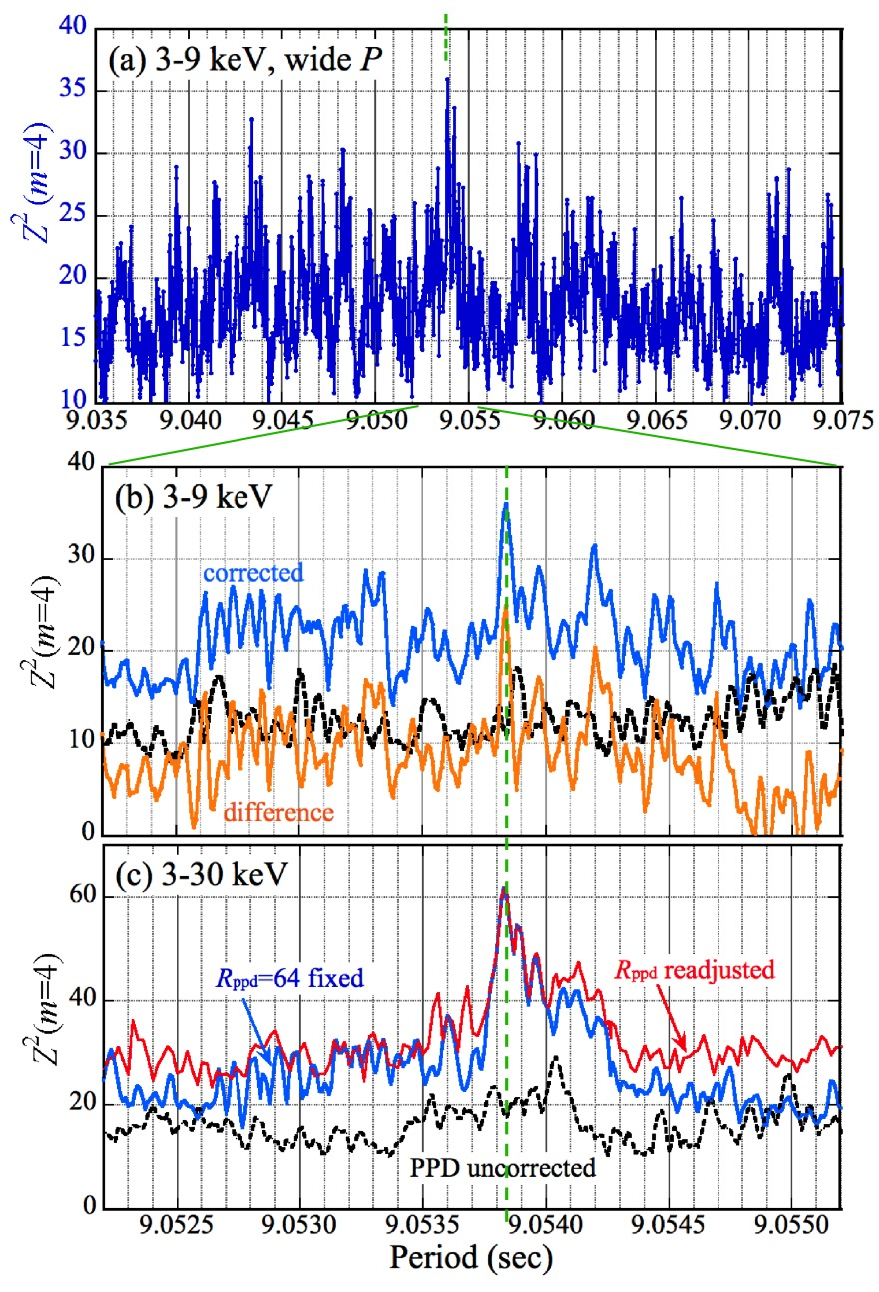}
}
\caption{
(a) A PPD-corrected 3--9 keV PG with $m=4$, 
using $\Eb=10$ keV and  $\R=62$ deg keV$^{-1}$.
Like in Figure~\ref{fig:Paper_I_review}(b), the orbital parameters 
except $e$ are varied at each $P$.
A vertical green line indicates $\Pns$ of Equation~(\ref{eq:P_NuSTAR}).
(b) Details of (a) around the peak,
shown over the same period range as in Figure~\ref{fig:Paper_I_review}(b).
Dashed black line gives the PPD-uncorrected PG,
and orange is the difference between blue and black.
(c) The same as (b), but  in 3--30 keV.
The blue trace employs $\Eb=10.4$ keV and  $\R=64$.
The red curve shows a result when $\R$ is also  allowed to float
(with $\Eb$ still fixed) at each $P$,
from $-90$ to 90 with a step of 1.0.
}
\label{fig:NuSTAR_Pscan_PPD}
\end{figure}

To confirm the pulse recovery in soft X-rays,
we  created a 3--9 keV PG in Figure~\ref{fig:NuSTAR_Pscan_PPD}(a),
incorporating the same PPD correction,
and the orbital-parameter readjustment at each $P$
like in Figure~\ref{fig:Paper_I_review}(b).
Compared to that fiducial PG,
the trial period range was expanded  by a factor of 13.
Then, the highest significance with $\zz=36.22$ is found at $P\approx 9.054$ s, 
whereas other peaks are all $\zz<34$.
As detailed in Figure~\ref{fig:NuSTAR_Pscan_PPD}(b),
the peak is indeed right at $\Pns$.
As shown therein by an orange trace,
the PPD correction also maximizes the $\zz$ increment 
at $\Pns$, $\Delta \zz \approx 25$.
Within $\sim \pm1$ ms of $\Pns$,
this $\Delta \zz$  takes systematically positive values,
probably because the pulse power, once restored by the PPD correction,
is partially scattered  by  the observing window.

Figure~\ref{fig:NuSTAR_Pscan_PPD}(c) 
shows a 3--30 keV broadband PG,
produced in a similar way to (b), 
but using $\Eb=10.4$ keV and $\R=64$ deg keV$^{-1}$.
Again at $P=\Pns$,  the PG achieves the peak with  $\zz = 61.51$,
which coincides in height with the red peak in Figure~\ref{fig:NuSTAR_Rscan}(b).
However, the peak still remains lower than in Figure~\ref{fig:Paper_I_review}(b).
This is because the 3--10 keV interval, 
where the pulse fraction is intrinsically lower,
contains twice as many photons as the 10--30 keV range,
and because PPD-corrected 3--10 profile
(black in Figure~\ref{fig:NuSTAR_PulseProfiles_PPD}a) 
somewhat differs from that in 10--30 keV (Figure~\ref{fig:Paper_I_review}d).

The 3--30 keV PG with no PPD correction,
shown in black in Figure~\ref{fig:NuSTAR_Pscan_PPD}(c),
reveals no peaks at $\sim \Pns$,
because of the dominance of soft X-ray photons.
This naturally explains how the \NuS\ pulsation escaped 
the reconfirmation by \cite{Volkov21},
who started the timing analysis in the 3--20 keV energy interval.

\subsubsection{Statistical significance of the PPD effect}
\label{subsec:NuS_PPD_significance}

When  calculating Figure~\ref{fig:NuSTAR_Pscan_PPD}(a),
we fixed  $\Eb$ and $\R$  to the values  optimized at $\Pns$.
Therefore, this PG is biased toward $\Pns$,
and does not correctly reflect  the enlarged parameter space.
However, if  $\R$ is also  allowed to float at each $P$,
all periods examined  will become equivalent,
and fluctuations in $\zz$ at $P \neq \Pns$, 
which must  be now larger, will provide  a reference with 
which we can evaluate the statistical significance of the peak at $P =  \Pns$.
(Since $\Eb=10$ keV is outside the energy range,
we can  fix it.)
In Appendix B, we carried out this attempt,
and found that the $\zz$ increase in 3--9 keV at $\Pns$,
through the PPD correction (Figure~\ref{fig:NuSTAR_Pscan_PPD}b),
has a false chance probability of $\Pch \sim 5\%$.

This  $\Pch$ in 3--9 keV is rather loose,  but it changes
when  the same evaluation is conducted in  the 3--30 keV broadband.
We obtain a very low chance probability of $\Pch \sim 0.004\%$ (Appendix B),
for a value of $\zz \geq 61.51$ to appear  
at $\Pns$ (Figure~\ref{fig:NuSTAR_Pscan_PPD}c)
via the PPD and orbital corrections,
in which the parameters except $e$ are all readjusted at each $P$.
Table~\ref{tbl:significance} summarizes these results,
together with those derived later from the \ASCA\ data.
Thus, the PPD effect,
operating below 10 keV of the  \NuS\ data,
is considered real (see \S~\ref{subsec:Pns_reinforce}).

For reference, the thin red line in Figure~\ref{fig:NuSTAR_Pscan_PPD}(c)
is  a  3--30 keV PG obtained by  allowing  $\R$ to float at each $P$.
(The  period search step of  $20~\mu$ s is recovered here.)
Thus, the effect of the enlarged parameter space is 
relatively limited in this energy range,
typically by $\Delta \zz \lesssim 10$,
which is much smaller than the systematic increase at $\Pns$,
$\Delta \zz \approx 42$.

\subsubsection{Constraints on the orbital parameters}
\label{subsub:NuS_soft_orbital}

\begin{figure}
\centerline{
\includegraphics[width=9cm]{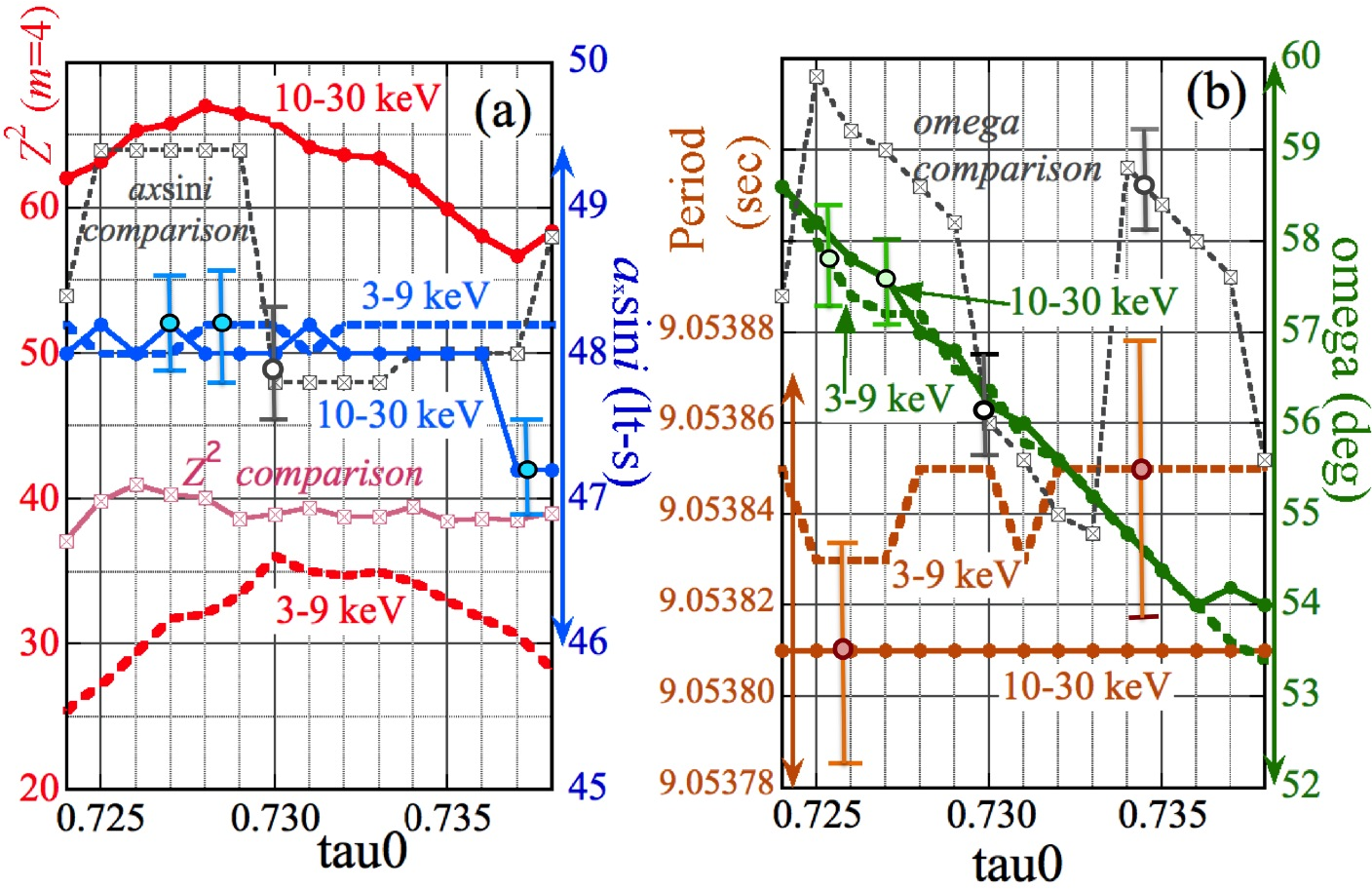}
}
\caption{Comparison of the orbital constraints in
10--30 keV (solid lines) and 3--9 keV 
(dashed lines;  $\Eb=10$ keV and $\R=62$ fixed).
The values of $P$, $\ax$, and $\omega$,
which maximize  $\zz$,  are shown as a function of $\tau_0$,
together with $\zz$.
The ranges over which 
these quantities are scanned are indicated by arrows along the ordinates.
The lines denoted as {\it comparison} are explained in text.
(a) $\zz$ (red) and $\ax$ (blue).
(b) $\omega$ (green) and $P$ (brown).
}
\label{fig:NuSTAR_tauscan_SoftHard}
\end{figure}

Returning to the 3--9 keV interval, 
a support (though not quantitative) to the reality of the PPD effect is provided
by Figure~\ref{fig:NuSTAR_tauscan_SoftHard} and 
Table~\ref{tbl:orbital_parameters},
where we  compare the orbital constraints  
in 10--30 keV (Paper I) and 3--9 keV,
derived without and with the PPD correction, respectively.
(Here, the orbital  parameters are allowed to vary over
wider ranges than in Figure~\ref{fig:NuSTAR_Pscan_PPD}, for presentation.)
The optimum values of $\omega$ (panel b) and $\ax$ (panel a),
derived in 3--9 keV using  $\Eb=10$ keV and $\R=62$,
are seen to depend on $\tau_0$ 
nearly in the same way as the 10--30 keV solution
which is free from the PPD disturbance.
The difference in $P$ by $\approx 40~\mu$ s is still within relative errors.
Thus, the 10--30 keV photons and the PPD-corrected 3--9 keV photons,
which are independent, yield nearly the same orbital constraints;
they are hence thought to represent the same phenomenon.

A contrasting case is shown in Figure~\ref{fig:NuSTAR_tauscan_SoftHard}
 by thin lines denoted as {\it comparison}.
They represent a typical  {\it false}  solution in 3--9 keV,
which we encountered in the calculation of Appendix B.
Characterized by  $P=10.5285$ s and $R=76$ deg keV$^{-1}$,
it gives $\zz=40.6$,
and its values of $\ax$ and $\omega$
agree with the 10--30 keV results,
if evaluated at a single point of 
$\tau_0 \approx 0.730$.
Nevertheless, if regarded as a function of $\tau_0$,
this false solution behaves in much different ways from the fiducial one.

Through the reanalysis of the \NuS\ data,
we have thus arrived at a scenario
that the object is pulsating even in energies below 10 keV,
but the pulse phase suffers the PPD effect expressed by Equation~(\ref{eq:PPD}).
This result reinforces the pulse credibility
(see \S~\ref{subsec:Pns_reinforce}),
and provides an important step forward  in solving the issue (ii).
It is  also suggested
that the search for pulsations in \LS\ can be carried out
using rich archival data below $\sim 10$ keV,
where  previous attempts were hampered presumably by the PPD perturbation 
which was not noticed then.

\begin{table*}
\begin{center}
\caption{A summary of the statistical significance of the pulsation.}
\label{tbl:significance}
\vspace*{-3mm}
\begin{tabular}{lccccccccccc}
\hline
      & Energy &\multicolumn{2}{c}{Corrections} & Period search&  $P$ 
                                        & $m$ &$Z_m^2$ & ${\Pch}^{a)}$ & Method$^{\,b)}$& Citation\\
         & (keV)    &  orbit & PPD  & range (sec)  & (sec)  &   & & (\%)\\
 \hline
  \hline
\multicolumn{4}{l}{Suzaku (2007)}\\
& 10--30&  no & no & 1--100  & $8.96 ^{\,c)}$
                        & 1& ---$^{d)}$   &0.15 &  MC    & Paper I \\
\hline
\multicolumn{4}{l}{NuSTAR (2016)}\\
& 10--30&  no & no & 7--11  & $9.046^{\,c)}$
                        & 1 & ---$^{d)}$   & 3.1 &  MC    & Paper I \\
& 10--30&  yes & yes & 9.025--9.065   & $\Pns$ [Eq.(\ref{eq:P_NuSTAR})] 
                        & 4 & 66.87  & 7$\,\dagger$ &  MC    & Paper I \\
&  3--9    &  yes & yes & $\sim \Pns$  & $\Pns$ [Eq.(\ref{eq:P_NuSTAR})]  
                        & 4 & 36.22  & 5$\,\dagger$  & Control &  App.B, \S~\ref{subsec:NuS_PPD_significance} \\
&  3--30  &  yes & yes & $\sim \Pns$  &$\Pns$ [Eq.(\ref{eq:P_NuSTAR})]  
                         & 4 & 61.51  &$0.004\,\dagger$ & Control & App.B, \S~\ref{subsec:NuS_PPD_significance} \\
\hline
\multicolumn{4}{l}{ASCA GIS (1999)}\\
~Stage 1 &  6--12  &  no & no  & 8.2--9.2  & $P_0$ [Eq.(\ref{eq:P0})]$^{\,c)}$ 
                           & 2 & 25.1  & 16& Analytic &  \S~\ref{subsubsec:step1}\\
               &  6--12  &  no & no  & 8.82--8.90  & $P_0$ [Eq.(\ref{eq:P0})]$^{\,c)}$ 
                            & 2 & 25.1  & 1.2 & Analytic &  \S~\ref{subsubsec:step1}\\
~Stage 3 &  2.8--12  & (yes)$^{e)}$& yes & 8.82--8.90  & $P'_1$  [Eq.(\ref{eq:Pdot2}a)]$^{\,c)}$ 
                            & 4 & 53.6  &  $0.2\,\dagger$  & Control & App.C, \S~\ref{subsubsec:step3}\\
 \hline
\end{tabular}
\end{center}
\begin{itemize}
\setlength{\itemsep}{-0.3mm}
 \item[$^{a)}$]: The items with $\dagger$ are utilized in the final significance estimation in \S~\ref{subsec:Pns_reinforce}.
   \item[$^{b)}$]: The method used in the significance evaluation.
   ``MC'' means Monte-Carlo simulations, and ``Control'' means 
   a use of the same data over different period ranges.
 \item[$^{c)}$]: These are intermediate periods, 
  and are somewhat different from the final periods given Table~\ref{tbl:orbital_parameters}.
 \item[$^{d)}$]: Not shown because summed values of $Z_m^2$ are used.
 \item[$^{e)}$]: The orbital Doppler effects are emulated by a constant $\dot P$.
\end{itemize}
\end{table*}

\section{ANALYSIS OF THE \ASCA\ GIS  DATA}
\label{sec:GIS_analysis}

Based on the above prospect,
we analyze the 0.7--12 keV data  of \LS\  acquired with \ASCA\  in 1999.
The aims are to strengthen the evidence for the $\sim 9$ s pulsation,
and to examine whether the PPD effect 
noticed in \S~\ref{sec:NuSTAR_reanalysis} is present or not.

\subsection{Observation}
\label{subsec:observation}

LS 5039 was observed once with \ASCA\
\citep{Tanaka94}, on 1999 October 4,
which is 7.94 yrs and 16.92 yrs before the \Su\ and \NuS\ observations, respectively.
The  gross exposure  (the total data span) is $T=63$ ks, 
or  $0.19P_{\rm orb}$,
whereas the net exposure is about 45\% of that.
We utilize data from the 
Gas Imaging Spectrometer \citep[GIS:][]{GIS1,GIS2}, 
placed at the focal planes of the X-ray Telescope \citep{XRT}.
The GIS covers a  0.7--12 keV energy range
with  moderate angular and energy resolutions.
The time resolution is 61 or 488 $\mu$s,
depending on the  telemetry rate.
\cite{Kishishita09} used these GIS data 
for spectral and photometric studies,
but not for timing analyses.

The data were processed in a standard way. 
From the identical pair of focal-plane detectors, GIS2 and GIS3,
we extracted on-source events  over a circular region 
of radius $5'$  centered on the source,
and co-added the GIS2 and GIS3 events.
The arrival time of each event was converted to the barycentric value.
Our timing analysis is performed on these photons,
without subtracting the background
which amounts to $20\%$ and $40\%$ of the total events
at 2--3 keV and 10--12 keV, respectively.

The first photon in the data was recorded at MJD 51455.529. 
It translates to an initial orbital phase of
\begin{equation}
\tau_0=0.35~~{\rm or} ~~0.44
\label{eq:ASCA_orbital_phase}
\end{equation}
based on the \Su\ or \NuS\ ephemeris (Paper I),  respectively.
The discrepancy between the two values 
reflects the issue (iii) in \S~\ref{sec:intro}.
Alternatively, optical observations suggest $ \tau_0=0.30$
 \citep{Kishishita09}.
Thus, we must treat $\tau_0$ as having a considerable uncertainty.
Regardless of this, 
the final phase of the GIS data is $\tau_0+0.19$.
Across the observation,
the background-inclusive GIS2 +GIS3 count rate 
in 1--12 keV gradually increased 
from $\approx 0.2$  to $\approx 0.3$ cts s$^{-1}$.
The X-ray light curve of LS 5039 has a good
reproducibility  \citep{Kishishita09, Takahashi09,Yoneda23},
and this brightening agrees with what is  expected 
for the estimated orbital phase.

We conducted a coarse spectral study
by subtracting an appropriate off-source background.
The derived 1--12 keV spectrum is featureless,
and can be fit adequately by an absorbed power-law model.
The  photon index is $1.6 \pm 0.1$, 
the absorbing column is $N_{\rm H}=(6.4 \pm 1.5)\times 10^{21}$ cm$^{-2}$, 
and the absorption-removed 2--10 keV flux 
is $(9.7 \pm 0.4) \times 10^{-12}$ erg cm$^{-2}$ s$^{-1}$.
These are typical of \LS,
and agree very well with those in \cite{Kishishita09}.

\subsection{Timing Analyses of the GIS data}
\label{subsec:rGIS_timing}

The pulse periods from the \Su\ and \NuS\ observations,
given in Table~\ref{tbl:orbital_parameters},
define an average pulse-period change rate as
$\dot P = 3.4 \times 10^{-10}$ s s$^{-1}$ (Paper I). 
A back extrapolation assuming this $\dot P$ predicts
the period  at the \ASCA\ observation as
\begin{equation}
P=8.871~~(8.826-8.900) ~~\rm s.
\label{eq:ASCA_P_prediction}
\end{equation}
Here, the upper and lower bounds assume 
that the actual $\dot P$ was 
1.5 and 2/3 times the above nominal value, respectively.
This period tolerance is considered  wide enough.
In fact, when the pulse-period evolution is expressed
as a quadratic function of time $t$ as
\[
P(t) = {\rm constant}+\dot P t + \tfrac{1}{2} \ddot P\, t^2,
\]
the upper and lower bounds  assumed in Equation~(\ref{eq:ASCA_P_prediction})
imply period-change time scales of
\[
\dot P/\ddot P \approx \pm 20~\rm{yrs}
\]
which matches the overall observational span of the present study.
In addition, the allowance is wide enough to accommodate glitches,
in which $P$ would change, in fraction,
by $\sim 10^{-5}$ at most \citep{glitches};
this is far smaller 
than the allowance in  Equation~(\ref{eq:ASCA_P_prediction}).

When searching the GIS data for the pulsation,
the parameter space to be examined is huge \citep[see also][]{Volkov21},
because of the uncertainty in Equation~(\ref{eq:ASCA_P_prediction}),
the \Su\ vs. \NuS\ ambiguity in the orbital solutions,
the need to consider the PPD effects,
and the choice  of energy ranges.
At the same time, the search grids must be fine enough,
because a periodicity in the signal is often  accompanied by 
many side lobes arising via  interferences 
with the quasi-periodic data gaps.
Thus, instead of searching at once the whole  parameter space
using fine grids (spending unrealistic computational times),
we proceed in four ``Stages"
with progressive sophistication and complexity.
We can then narrow down the parameter space in stepwise ways,
and filter out false side lobes based on  their statistical and systematic behavior.
This strategy is the same as employed successfully in Paper I,
although the actually employed  stages are not the same.

\begin{figure}
\medskip
\centerline{
\includegraphics[width=8.7cm]{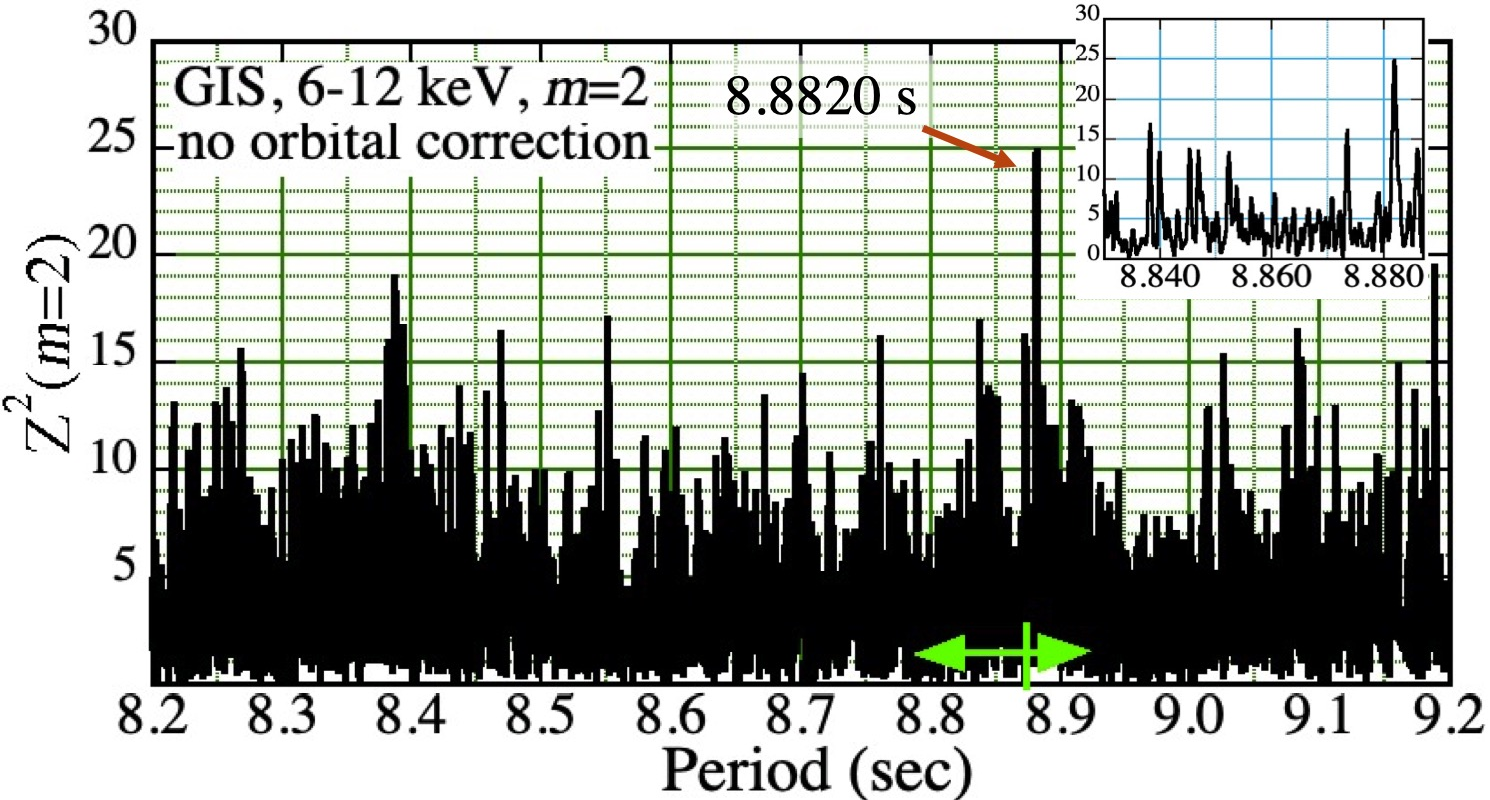}
}
\caption{
An $m=2$ PG from the 6--12 keV GIS data in Stage 1,
generated without orbital corrections.
The inset shows an expansion leftward of the peak at $P=P_0=8.8820$ s.
The green arrow represents Equation~(\ref{eq:ASCA_P_prediction}).
}
\label{fig:GIS_PG_m2_6-12_raw}
\end{figure}

\subsubsection{Stage 1: Simple periodograms}
\label{subsubsec:step1}

As the 1st Stage,
we tentatively ignore the orbital Doppler effects,
and produce $Z_2^2$ PGs
over a period range of 
8.2--9.2 s,
which includes, and is some 14 times wider than,
the range of Equation~(\ref{eq:ASCA_P_prediction}).
The trial period is varied with a step of $\Delta P=100~\mu$s.
We use $m=2$, because the pulse profiles would still be
smeared by the orbital Doppler effects,
and  lack sharp features.
Keeping the upper energy boundary at 12 keV,
we  tested four values of $E_{\rm L}$; 5.5, 6.0, 6.5, and 7.0 keV.

As  in Figure~\ref{fig:GIS_PG_m2_6-12_raw},
a promising result was derived  when using the 6.0--12 keV interval;
the PG reveals a clear peak, with  $Z_2^2 =25.1$,  
at  a period of
\begin{equation}
P= 8.8820~{\rm s}  \equiv P_0
\label{eq:P0}
\end{equation}
which is within the tolerance  of Equation~(\ref{eq:ASCA_P_prediction}).
Since $Z_2^2$ obeys a chi-square distribution of 4 degrees of freedom,
the chance probability for a value of  $Z_2^2 \geq 25.1$
to appear in a single trial is $\Pch^{(1)} = 4.8 \times 10^{-5}$.
Given the data span of $T=63$ ks,
the examined period range contains about 
$T/8.2-T/9.2 \approx 835$ independent Fourier waves.
The post-trial probability hence becomes
$\Pch \approx 835\;\Pch^{(1)} \approx 4.0\%$.
After multiplying this by four,
the number of $E_{\rm L}$ tested,
we obtain $\Pch = 16\%$
which is  given in Table~\ref{tbl:significance} 
together with those from \NuS.

The above result, 16\%, is relatively secure,
but not constraining.
As already justified,
we may  consider only those trials which fall in the 
uncertainty range of Equation~(\ref{eq:ASCA_P_prediction}).
Then, ${\Pch}$ reduces to  $\approx  1.2\%$ (Table~\ref{tbl:significance}),
which is sufficiently small.
Thus,  $P_0$ is regarded as a start point of
our \ASCA\ timing analysis.

When  $E_{\rm L}>6.0$  keV is used,
the PG peak at $P_0$ diminishes,
presumably because the photons would be too few.
The peak also decreases
when $E_{\rm L}$ is lowered from 6.0 keV,
in spite of an obvious increase in the signal photon number.
This suggests that the GIS data  are also affected,
at lower energies, by the PPD perturbation.

\subsubsection{Stage 2: Analysis considering $\dot P$}
\label{subsubsec:step2}

Although we have obtained reasonable evidence for a periodicity
that is consistent with those from \Su\ and \NuS,
we still need to perform two timing corrections
before  concluding on the pulse detection.
One is obviously that for the orbital delays,
and the other is that for the PPD effect as suggested above.
The former, though well formulated,
involves multiple parameters ($e,  \ax,  \omega$, and $ \tau_0$)
each with residual uncertainties (Paper I).
The latter, on the other hand, has only two parameters ($\Eb$ and $\R$),
but the formalism is only poorly established.
We also need to decide 
whether to conduct the two corrections simultaneously or in series,
and if the latter, which should be the first.

Fortunately, the orbital phase of Equation~(\ref{eq:ASCA_orbital_phase})
is such that the line-of-sight velocity  of the source
varies rather linearly with time $t$ (see \S~\ref{subsec:summary});
that is,  the pulse period changes  with an approximately constant rate of
\begin{equation}
\dot P_7 = 0.92 \pm 0.15,
\label{eq:Pdot}
\end{equation}
where $\dot P_7 $ represents $\dot P$ in units of $10^{-7} {~\rm s~s}^{-1}$.
The error includes  both the uncertainty of $\tau_0$ in Equation~(\ref{eq:ASCA_orbital_phase}),
and the changes during the observation.
We  hence remove first the orbital effects approximately, 
by considering $\dot P$ in the epoch-folding procedure,
in place  of the full orbital corrections.
The number of parameters then reduces from four to one,
which is  a big advantage.
We also switch from $m=2$ to $m=4$ (Paper I)
because fine structures will emerge in the folded pulse profiles.

\begin{figure}
\centerline{
\includegraphics[width=7cm]{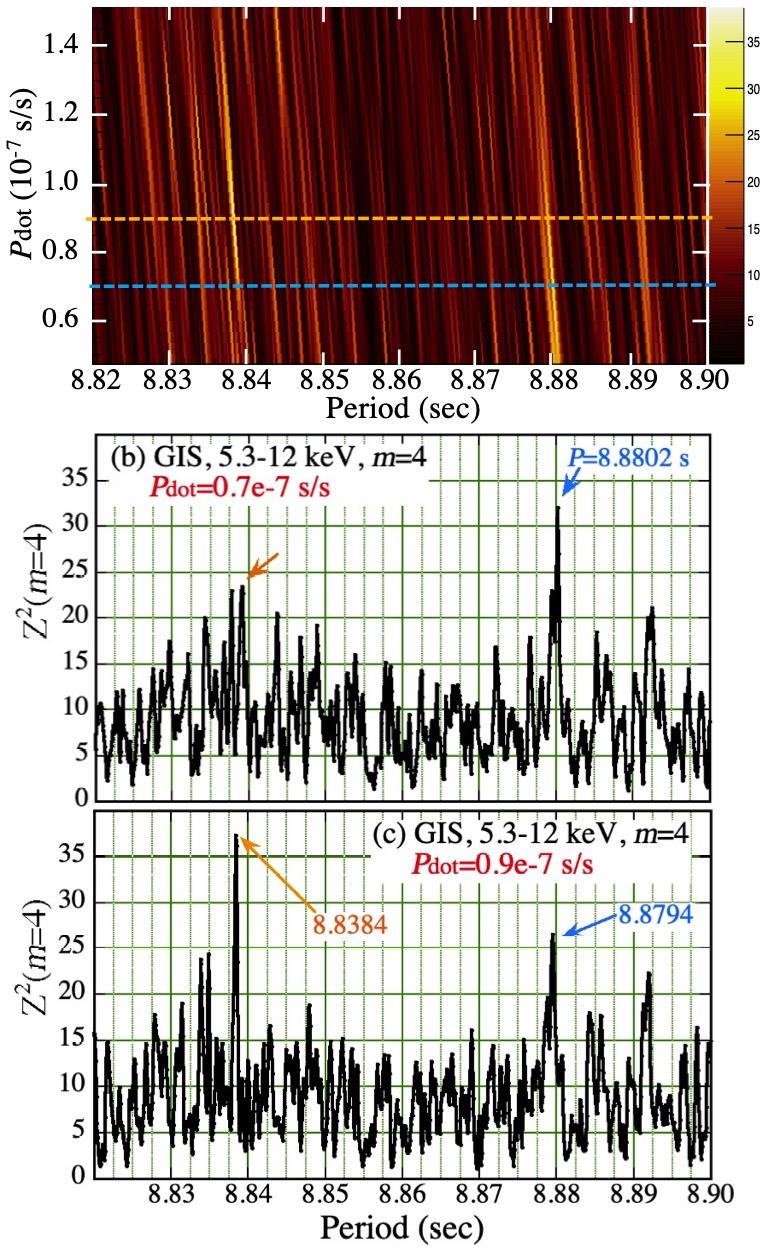}
}
\caption{Stage 2 results from the GIS:
epoch-folding  of the 5.3--12 keV GIS data with $m=4$,
incorporating $\dot P$.
(a) A color map of $Z_4^2$ on a $(P, \dot P)$ plane.
(b) A cross section of (a) at $\dot P_7= 0.7$ (dashed cyan line).
(c) The same, but at $\dot P_7 = 0.9$ (dashed orange line).
}
\label{fig:GIS_PGs_w_Pdot}
\end{figure}

As  our Stage 2 attempt, we repeated the PG calculation with the GIS data,
this time incorporating $\dot P$,
to obtain  $Z_4^2$  on a $(P, \dot P)$ plane.
We scanned $P$ over 8.82--8.90 s
which is comparable to that of Eaquation~(\ref{eq:ASCA_P_prediction}),
with an increment of $\Delta P=73.2~\mu$s,
and $\dot P_7$  from 0.5 to 1.5 with an increment  of 
$\Delta P_7=0.0313$.
Like in Stage 1, we tested several values of $E_{\rm L}$,
to compromise the signal statistics and the PPD disturbances.
Then, $E_{\rm L}=5.3$ keV was found to generally maximize $\zz$,
rather than the 6.0 keV threshold favored in Stage 1.

The result derived from the 5.3--12 keV interval (total 859 photons)  
is presented in Figure~\ref{fig:GIS_PGs_w_Pdot}(a).
Many yellow straight lines run in parallel,
with an inclination of $\approx  T/2$ against the abscissa.
Particularly bright are two of them, which are expressed on the plane as
\begin{equation}
\begin{split}
P ({\rm s}) &= 8.8820 - \tfrac{1}{2} T {\dot P} = 8.8820 - 0.0032 {\dot P_7}   \equiv P_1\\
P ({\rm s})  &= 8.8412 - \tfrac{1}{2} T {\dot P} =8.8412 - 0.0032 {\dot P_7}  \equiv P_2~.
\end{split}
\label{eq:GIS_two_solutions}
\end{equation}
They are  identified in panels (b) and (c) of Figure~\ref{fig:GIS_PGs_w_Pdot},
which provide two horizontal cuts across panel (a).
Of these, $P_1$  evidently connects to the $P_0$ peak 
in Figure~\ref{fig:GIS_PG_m2_6-12_raw}.
The orbital correction is thus  working at least on the $P_1$ branch,
because it becomes highest at $\dot P \sim 0.7$ as expected,
reaching $Z_4^2=32.1$ which  exceeds the value  ($Z_4^2=22.3$) at $\dot P=0$.
The other branch, denoted as $P_2$, becomes higher than $P_1$ at $\dot P \sim 0.9$,
whereas it gets weaker toward $\dot P \rightarrow 0$,
connecting to a weak counterpart  in the inset to  Figure~\ref{fig:GIS_PG_m2_6-12_raw}.

Below, we focus on $P_1$ and $P_2$.
Either of them is actually a family of parallel lines,
with a separation in $P$  by 2--15 ms.
This fine structure is  created  presumably when an enhanced power in the data,
from either the pulsation or noise fluctuations,
is split by the observing window of \ASCA\
which is a near-Earth satellite.
Its data are subject to data gaps,
 roughly synchronized
with the spacecraft's orbital period $P_{\rm sc} \approx 5.5$ ks.
Hereafter, we collectively call the family near $P_1$ ``Solution 1'' (or SOL-1),
and that near  $P_2$ ``Solution 2'' (or SOL-2).

\subsubsection{Stage 3: PPD corrections}
\label{subsubsec:step3}

\begin{figure}
\includegraphics[width=7.9cm]{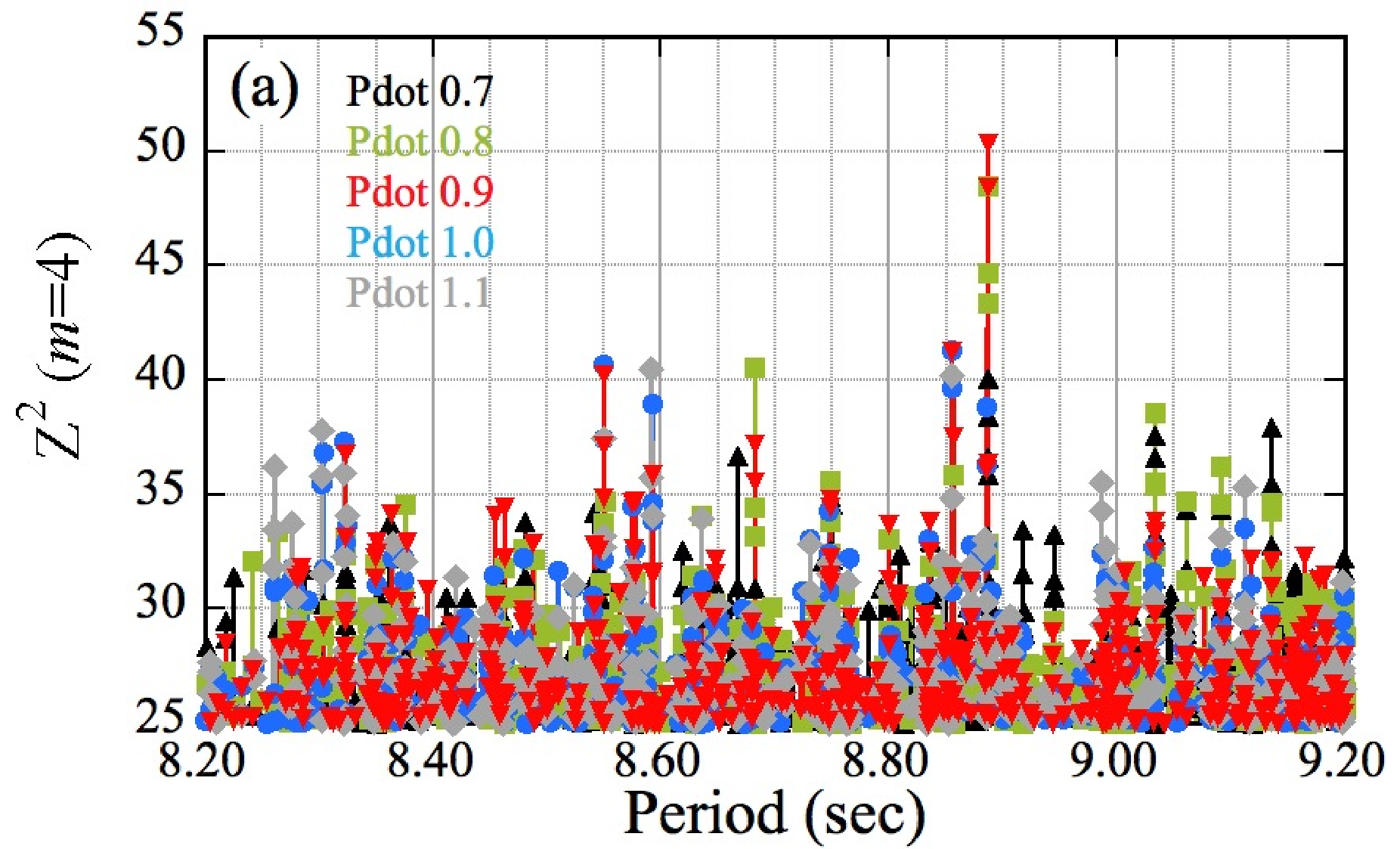}
\includegraphics[width=8.6cm]{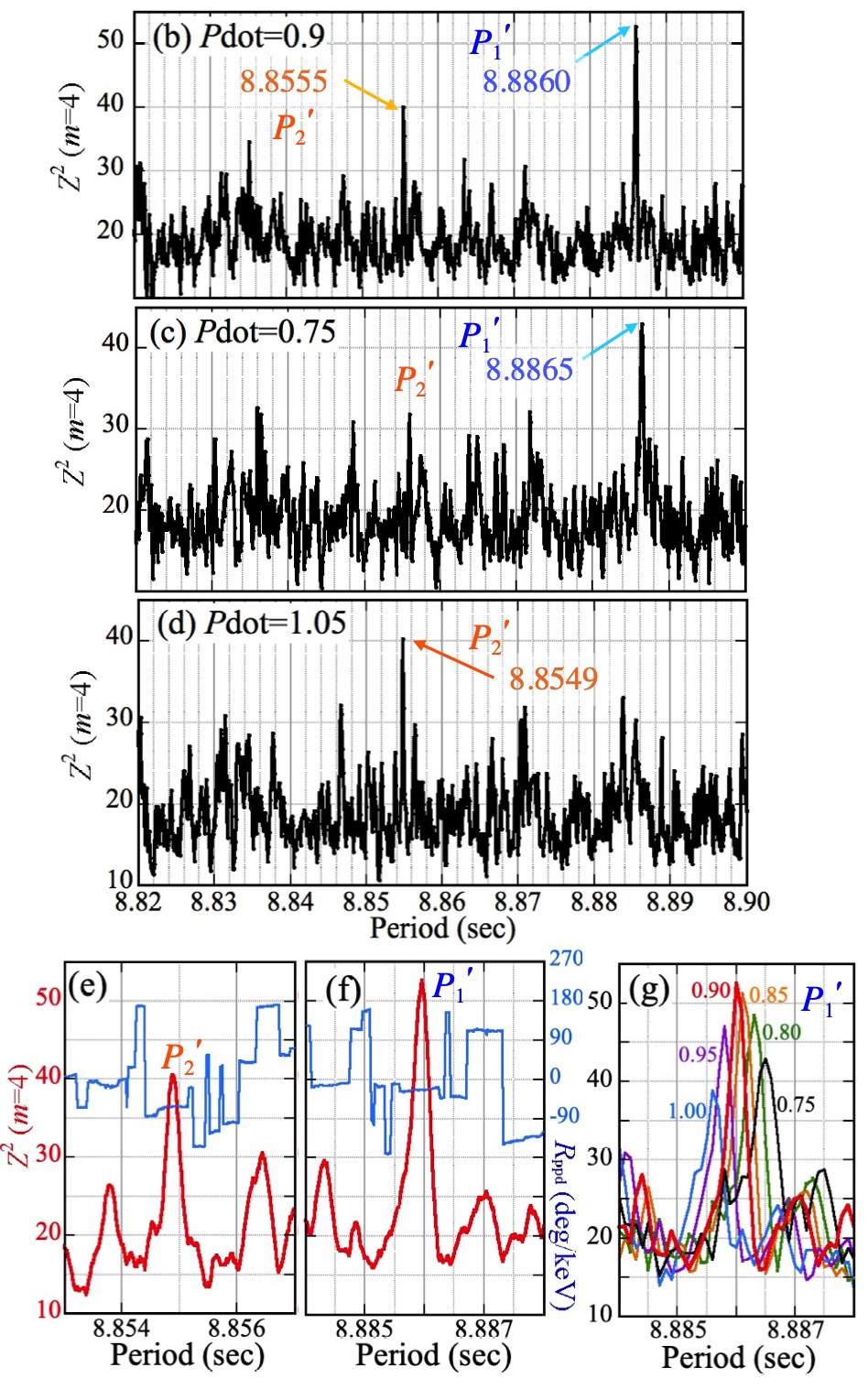}
\caption{Stage 3 results from the GIS.
(a) PGs with $m=4$ in  2.8--12 keV,
calculated over a broad period range with $\Delta P=100~\mu$s,
using the PPD correction in which $\R$ is 
optimized at each $P$ (see text).
Black, green, red, blue, and gray assume 
$\dot P_7=0.7, 0.8, 0.9, 1.0$ and 1.1, respectively.
(b), (c), and (d) show results over the limited period rage,
using $\Delta P=20~\mu$ s,
for $\dot P_7=0.9, 0.75,$ and 1.05, respectively.
(e) and (f) give details of the $P_2'$ peak in (d) 
and the $P_1'$ peak in (b), respectively,
where the blue trace presents the optimum $\R$.
(g) A superposition of the $P_1'$ peaks
for six  values of $\dot P$ (given in color) 
which cover the range of Equation~(\ref{eq:Pdot}).}
\label{fig:GIS_Rscan}
\end{figure}

Now that the orbital effect is  successfully
approximated by $\dot P$,
we proceed to Stage 3, namely, 
incorporation of the PPD correction using Equation~(\ref{eq:PPD}). 
Considering that the correction for \NuS\ 
worked down to the instrumental lower bound of  3.0 keV,
the analysis utilizes  $E_{\rm L} =2.8$ keV, 
which is justified later.
Then, we specify a value of $\dot P$ 
in the range of  Equation~(\ref{eq:Pdot}),
and change $P$ over the broad  interval 
used in Stage 1 (Figure~\ref{fig:GIS_PG_m2_6-12_raw})
with $\Delta P =100~\mu$s.
At  each $P$, we scan $\R$
with a step of 1.0 deg keV$^{-1}$  like in the case of \NuS, 
using a wider allowance 
from $-180$ to $180$ deg keV$^{-1}$, 
and register the maximum $\zz$ 
together with the associated $\R$.
Tentatively, $\Eb$ is fixed  at 10.0 keV.

The above procedure was carried out for 
five values of $\dot P_7$, from 0.7 to 1.1 with a step of 0.1.
The obtained five PGs are superposed in Figure~\ref{fig:GIS_Rscan}(a).
A dominant peak reaching $\zz \approx  52$ has
emerged at $P \approx  8.89$ s,
which favors $\dot P_7=0.9$ (red).
For reference, we tentatively expanded the period search range
further to 7.0--11.0 s, as used in Paper I for the initial study of the \NuS\ data,
but the peak at 8.89 s still remained the highest,
with the next one having $\zz \approx  49.1$.

In the same way as in Stage 2, 
the period search range is hereafter limited to 8.82--8.90 s,
which accommodates the dominant peak.
We re-generated PGs over this range, 
with a finer step of $\Delta P =20~\mu$s.
The results are shown in panels (b), (c), and (d) of Figure~\ref{fig:GIS_Rscan},
which employ  $\dot P_7=0.9$, 0.75, and 1.05, respectively.
The dominant peak in (a) is reproduced in (b) and (c),
at a period of $P \approx 8.886$ s, which we hereafter denote as $P_1'$.
Although it differs by $\Delta P \approx 6$ ms
from $P_1$ itself ($P_1=8.8791$ s for $\dot P_7=0.9$),
it clearly belongs to the SOL-1 family, 
because  we find $(1/P_1 -1/P_1')^{-1} = 2.07 P_{\rm sc}$
which indicates interference by the observing window.
The secondary peak in (a) becomes highest in (d), 
and seen at $P \approx 8.855$ s $\equiv P_2'$.
This $P_2'$ peak belongs to the SOL-2 family,
because  $(1/P_2 -1/P_2')^{-1} = 0.84 P_{\rm sc}$ holds,

Regions around the $P_2'$ peak in Figure~\ref{fig:GIS_Rscan}(d)
and around the $P_1'$ peak in Figure~\ref{fig:GIS_Rscan}(b) 
are expanded  in  panels (e) and (f), respectively.
There,  the optimum value of $\R$ is shown in blue.
Thus, $P_1'$ and $P_2'$ demand
$\R \approx -23$ and $\approx -63$ deg keV$^{-1}$, respectively.
Although the height of $P_2'$  reaches a maximum of $\zz =  44.7$
when the involved parameters are trimmed,
it  remains considerably lower than that of  $P_1'$.

From these examinations,
we regard SOL-1 as  the  more likely solution family,
among which  $P_1'$ is  the best pulse-period candidate 
under the Stage 3 formalism.
The $P_1'$ peak in Figure~\ref{fig:GIS_Rscan}(f) 
is unlikely to be caused by statistical fluctuations, 
because its  chance  probably is sufficiently low,
$\Pch \approx 0.2\%$ (Table~\ref{tbl:significance}), 
as  evaluated in Appendix C.
We hereafter concentrate on  $P_1'$.

The $P_1'$ peak in Figure~\ref{fig:GIS_Rscan}(f) is  very sharp, 
because these PGs are calculated 
each for a fixed value of $\dot P$.
To reflect  the obvious error propagation from $\dot P$ to $P$,
Figure~\ref{fig:GIS_Rscan}(g) superposes the $P_1'$ peaks from five values of $\dot P$.
The composite peak is  broader not only than that in Figure~\ref{fig:GIS_Rscan}(f),
but also than those  in Figure~\ref{fig:Paper_I_review} by several times,
due to the shorter data span of \ASCA.

Figure~\ref{fig:GIS_Rscan}(g) displayed 
$\zz$ using $P$ as an independent variable,
$\dot P$ as a secondary parameter,
and $\R$ as an implicit parameter allowed to vary over a certain range.
By interchanging the roles of $P$ and $\R$ (but keeping that of $\dot P$),
we obtain Figure~\ref{fig:GIS_Rscan_Routput},
which is similar to  Figure~\ref{fig:NuSTAR_Rscan}.
Thus,  the 2.8--12 keV GIS data constrain $\R$ very tightly.
Using these results, and further  trimming $\Eb$, 
the maximum pulse significance of $Z_4^2=53.6$ has been 
obtained under a condition of
\begin{subequations}
\begin{align}
&P'_1=8.8860 \pm 0.0003~{\rm s}\\
& \dot P_7 = 0.88 \pm 0.10
\end{align}
 \label{eq:Pdot2}
\end{subequations}
together with  $\R=-23.4^{+2.6}_{-3.7}$ deg keV$^{-1}$
and $\Eb =10.0 ^{+2.0}_{-2.3}$ keV.
This $\Eb$ is very close to that found with \NuS,
{but subject to a larger error,
because it is near the upper threshold of the GIS.
As the largest difference from \NuS,
 $\R$ has the opposite sign.

The GIS data for the first time  provide
the pulse information of \LS\ below $\sim 3$ keV.
We  hence varied $E_{\rm L}$ of the analysis,
and found that the employed 2.8 keV  gives the highest $\zz$.
While the $\zz$ decrease toward higher $E_{\rm L}$ can be
explained by a decrease of the photon number,
that from $E_{\rm L} = 2.8$ keV downward is not trivial.
So, we tried the pulse search in 1.0--2.8 keV
incorporating the orbital and PPD corrections,
with $\Eb$ and $\R$ allowed to float.
However, no evidence of pulsation at $\approx P_{\rm fin}$ 
was found ($\zz<23 $) in this softest interval,
which contains 17\% more photons than the 2.8--12 keV band.
The pulse properties might change 
again at $\sim 2.8$ keV.

\begin{figure}
\centerline{
\includegraphics[width=8.2cm]{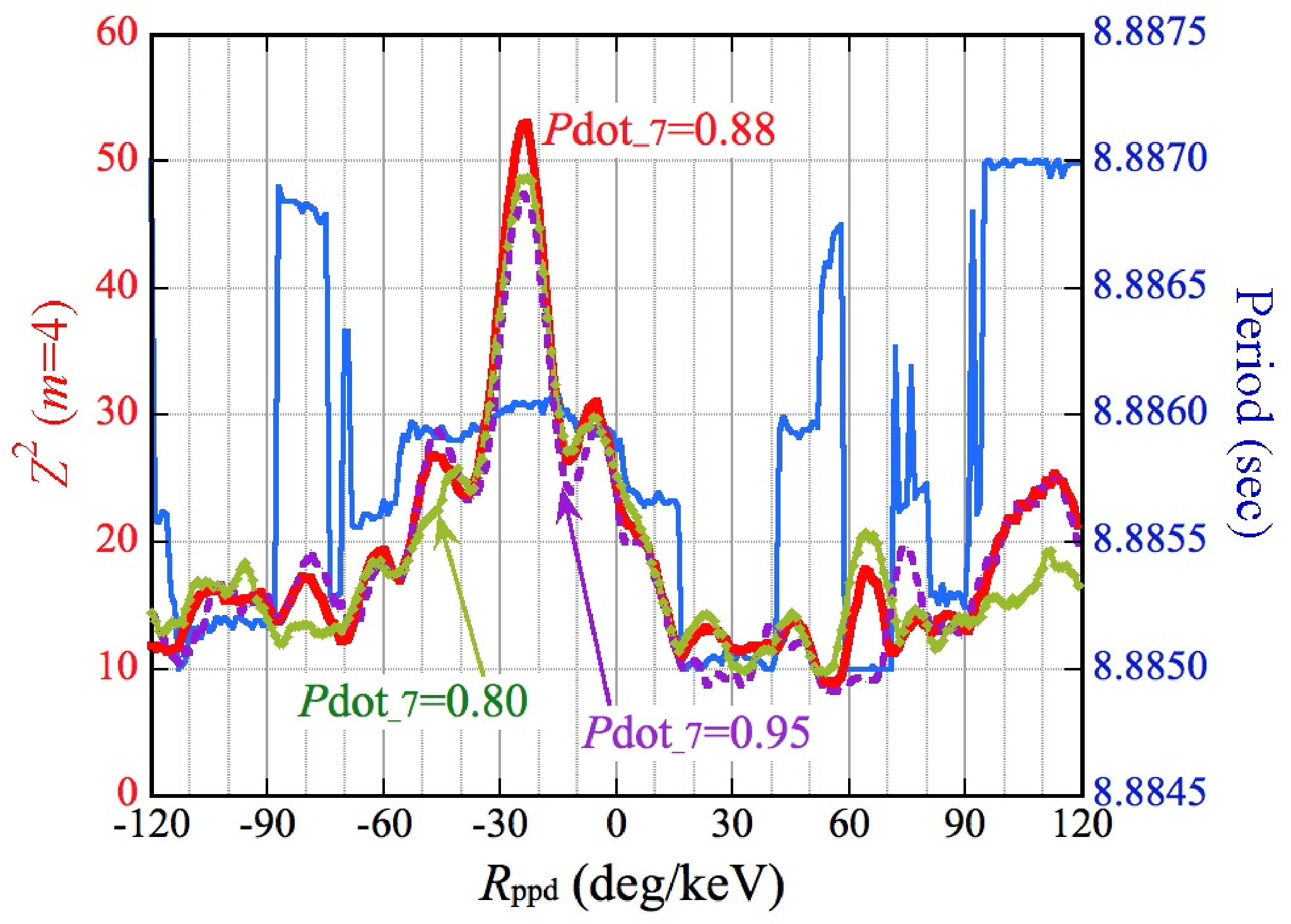}
}
\caption{The same as Figure~\ref{fig:NuSTAR_Rscan}(b),
but for the 2.8--12 keV  GIS data in Stage 3.
The $\R$ dependence of $\zz$ is shown for
three values of $\dot P_7$; 0.80 (green), 0.88 (red),
and 0.95 (dashed purple).
At each $\R$, $P$ is scanned over $8.8860 \pm 0.0015$ s,
and its optimum value  for $\dot P_7=0.88$ is presented in blue
(right ordinate).
In all cases, $\Eb=10$ keV is retained.
}
\label{fig:GIS_Rscan_Routput}
\end{figure}

To see whether the linear energy dependence 
assumed in Equation~(\ref{eq:PPD}) is appropriate,
Figure~\ref{fig:GIS_Rscan_Routput} was recalculated
for two separate energy bands, 2.8--5 keV and 5--12 keV,
both using $\dot P_7=0.88$.
The peak was clearly detected in both of them (though not shown),
at $\R=-20^{+13}_{-14}$ ($\zz=36.2$) in 2.8--5 keV,
and $\R=-22\pm 6$ ($\zz=22.4$) in 5--12 keV.
The agreement between the two  $\R$ values
gives a  support to Equation~(\ref{eq:PPD}).
The $\zz$ peak is higher but  broader in the softer band, 
which has twice more photons but a narrow energy span.

Thanks to the PPD correction,
the periodicity found in Stage 1 and Stage 2 has thus been 
confirmed, with high significance,
over the broad energy interval of 2.8--12 keV.
Furthermore,  $P_1'$  is regarded as  intrinsic to \LS,
because the optimum $\dot P$ specified by the data,
Equation~(\ref{eq:Pdot2}b), approximately agrees with  
that predicted by the orbital Doppler effect,
Equation~(\ref{eq:Pdot}).

\subsubsection{Stage 4: Orbital corrections }
\label{subsubsec:step4}

Although our Stage 3 analysis was successful,
the $\dot P$ approximation utilized there
must be finally replaced with 
the proper correction for the elliptical orbit.
This makes our final (4th) Stage.
Hence, $\dot P$  is now reset to 0
as the secular spin-down rate ($\dot P_7\approx  0.0034$) is negligible.
We retain $m=4$,  and the 2.8--12 keV energy range
together with the PPD correction,
where $\Eb=10.0$ keV and  $\R=-23.4$ deg keV$^{-1}$ are fixed for the moment.
Then, $\zz$ is calculated as a function of $P$
from 8.80 s to 8.94 s (with a step of $100\mu$s).
Like in  Figure~\ref{fig:Paper_I_review}(b) and  Figure~\ref{fig:NuSTAR_Pscan_PPD},
at  each $P$ we scan  $\ax$ from 46.5 to 53.5 lt-s (0.5 lt-s step),
$\omega$ from $53^\circ$ to $60^\circ$ ($0^\circ.5$ step),
and $\tau_0$ from 0.25 to 0.60 (0.002 step).
The scan ranges of $\ax$ and $\omega$  are wider 
than those used in \S~\ref{sec:NuSTAR_reanalysis}
(and the scan steps are coarser),
to accommodate both the \NuS\ and \Su\ solutions 
(see Table~\ref{tbl:orbital_parameters}).
Our keen interest is  whether the \ASCA\ data favor
either of the two orbital solutions, or any third one.

\begin{figure}
\centerline{
\includegraphics[width=8.2cm]{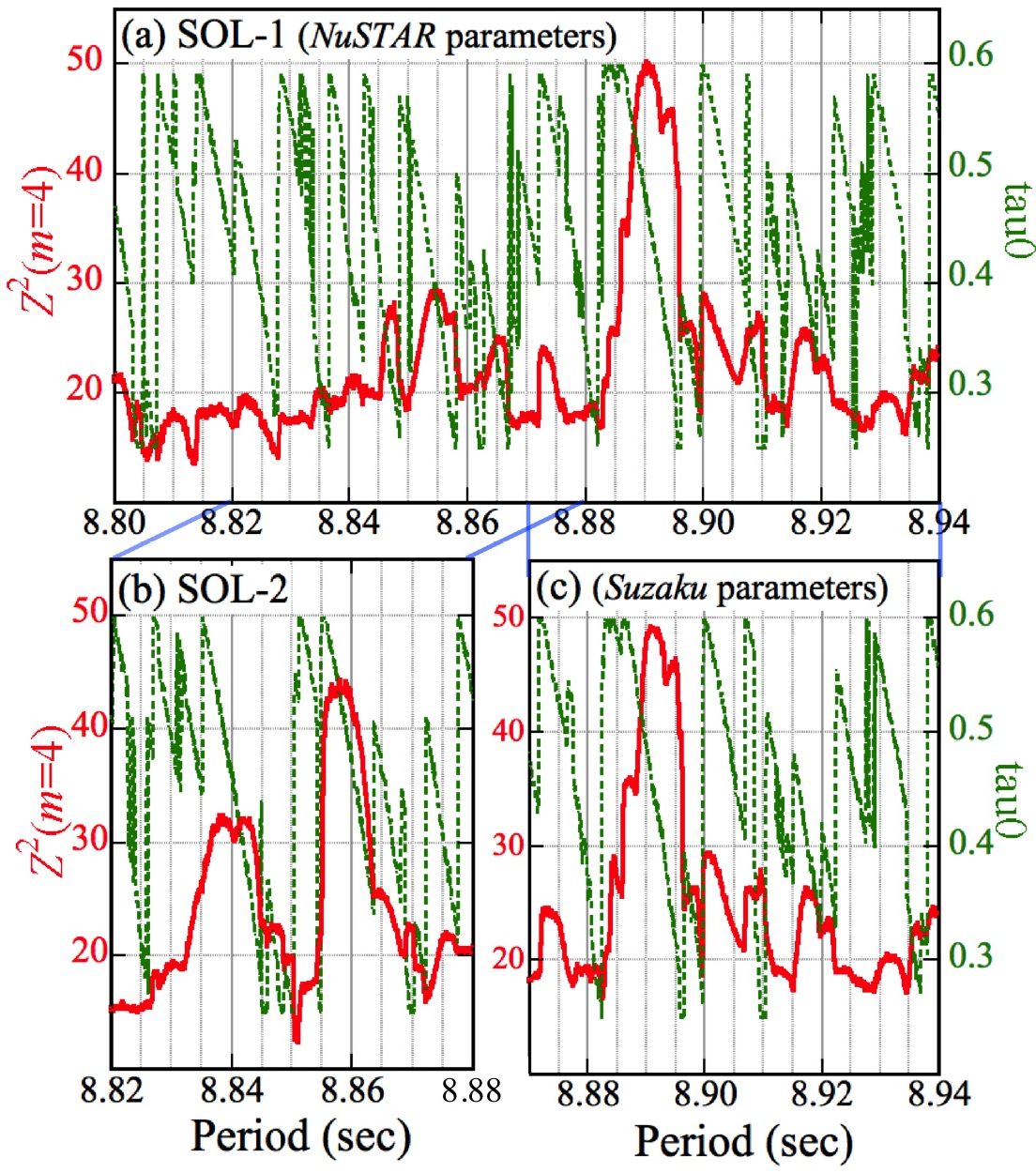}
}
\caption{The final (Stage 4) PGs from the 2.8--12 keV GIS data,
in which $\ax$, $\omega$, and $\tau_0$ are readjusted at each $P$, 
whereas the PPD parameters are fixed (see text).
The red and green traces show the maximum $Z_4^2$ and the optimum $\tau_0$, respectively.
(a) A result for SOL-1, assuming $e=0.306$, $\Eb=10.0$ keV, and $\R=-23.4$.
(b) A result for SOL-2, using $e=0.306$, $\Eb=10.5$ keV, and $\R=-62$.
(c) The same as (a), but  employing $e=0.278$ (from \Su),
together with $\Eb=10.1$ keV and $\R=-23.6$.
}
\label{fig:GIS_OrbCorr}
\end{figure}

When $e=0.306$ from the \NuS\ solution is assumed,
the PG shown in Figure~\ref{fig:GIS_OrbCorr}(a) was obtained.
It reveals a strong peak  at 
\begin{equation}
P \approx 8.891 ~{\rm s} \equiv P_{\rm fin},
\label{eq:P_final}
\end{equation}
which is about 30\% longer than the \Su\ plus \NuS\ prediction,
but still within the uncertainty of Equation~(\ref{eq:ASCA_P_prediction}).
As elucidated in \S~\ref{subsec:summary},
the difference $P_{\rm fin} - P_1' = 5$ ms agrees with the Doppler shift 
predicted by the ephemeris of Equation~(\ref{eq:ASCA_orbital_phase}).
Furthermore, as summarized in Table~\ref{tbl:orbital_parameters},
the derived $\ax$ and $\omega$ agree very well with those from \NuS, 
and $\tau_0=0.44 \pm 0.04$
is consistent with Equation~(\ref{eq:ASCA_orbital_phase}).

We repeated the same analysis,
but  using $e=0.278$ (Table~\ref{tbl:orbital_parameters})
 representing the \Su\ solution.
The PPD parameters were readjusted only slightly (Table ~\ref{tbl:orbital_parameters}).
The obtained result, shown in Figure~\ref{fig:GIS_OrbCorr}(c)
and Table~\ref{tbl:orbital_parameters},
is very similar to that in panel (a),
with insignificant differences in either the peak $\zz$ height 
or the pulse period $P_{\rm fin}$.
The optimum values of $\ax$ and $\omega$
have come to agree well  with the \Su\ solution,
whereas $\ax$ now disagrees with the \NuS\ value.

As seen from the \ASCA\ data,
the  two somewhat discrepant  orbital solutions
found with  \Su\ and \NuS\ thus degenerate, 
and their preference cannot be decided;
the issue (iii) in \S~\ref{sec:intro} remains unsolved.
This is because the \ASCA\ data cover only 0.19 orbital cycles of \LS,
whereas the \Su\ and \NuS\ data cover 1.5 and 1.0 cycles, respectively.
Nevertheless, the periodicity $P_{\rm fin}$ found with \ASCA\ satisfies
a few  important criteria; it is statistically significant (\S~\ref{subsubsec:step1}),
it  lies close to the \Su\ plus \ASCA\ extrapolation,
it  is consistent with the orbital Doppler effects,
and it shares the same $\Eb$ with the \NuS\ data.
We hence regard $P_{\rm fin}$ as the pulse period at the \ASCA\ observation,
which is unaffected by the \Su\ vs. \NuS\ ambiguity in the orbital solution.

\subsubsection{Additional remarks}
\label{subsubsec:GIS_additional}

To complement the Stage 4 analysis,
three remarks may be added.
First, the peaks in Figure~\ref{fig:GIS_OrbCorr}(a) and (c)  are considerably 
broader than the composite $P_1'$ peak in Figure~\ref{fig:GIS_Rscan}(g),
and by an order of magnitude than the \Su\ and \NuS\ results
(Figure~\ref{fig:Paper_I_review}, Figure~\ref{fig:NuSTAR_Pscan_PPD}).
This is  because  $P$ couples strongly with $\tau_0$ 
(dashed green trace in Figure~\ref{fig:GIS_OrbCorr}a)
under the very limited orbital coverage,
and $\tau_0$  is  allowed to vary  beyond  the interval 
where the $\dot P$ approximation works.

Next is how the orbital correction works on SOL-2.
This is shown in Figure~\ref{fig:GIS_OrbCorr}(b),
which was obtained  in the same way,
but employing $\Eb=10.5$ keV and $\R=-62 $ deg keV$^{-1}$.
The SOL-2 peak now appears at 8.858 s, or $P_2'+3$ ms,
and this increment  is consistent with the expected orbital effects.
However, the peak, with $Z_4^2=44.3$, is still significantly  lower  
than those in paneld (a) and  (c), $Z_4^2\approx 50$.
We reconfirm that $P_{\rm fin}$ of 
Equation~(\ref{eq:P_final}) is the best pulse period
at the barycenter of \LS.
The SOL-2 peak may be a side lobe of $P_{\rm fin}$.

Finally, we recalculated Figure~\ref{fig:GIS_OrbCorr}(a),
allowing $\R$ to float at each $P$.
Then, just like the relation between the blue and red traces 
in Figure~\ref{fig:NuSTAR_Pscan_PPD}(c),
the baseline  of $\zz$ was enhanced by $\Delta \zz \sim 10$,
but the region around  $P_{\rm fin}$ remained intact.
Allowing $\Eb$ to float gave no larger effects.
These results are not displayed.

\begin{table}
\caption{Pulse fraction in the 2.8--12 keV GIS data.}
\label{tbl:GIS_profile_eval}
\vspace*{-5mm}
\begin{center}
\begin{tabular}{lccccc}
\hline
Case & Pulse fraction & $\zz$  &$\Pch^{(1)}$ & $\chisq (\nu=19) $  \\
\hline
\hline
1        &$0.08 \pm 0.07$ & 12.3   & 0.13 &  2.12  \\
\hline
2        &$0.13\pm 0.07$  &  24.2  &$2.1 \times 10^{-3}$& 2.64 \\
\hline
3        &$0.21 \pm 0.08$ &  50.7 &$3.0 \times 10^{-8}$ & 3.66      \\
\hline
\end{tabular}
\end{center}
\begin{itemize}
\setlength{\itemsep}{-0.3mm}
  \item[Case 1]: No timing corrections (except  barycentric).
  \item[Case 2]: Case 1 plus the orbital correction  (Figure~\ref{fig:GIS_Pr_before_after}a).
  \item[Case 3]: Case 2 plus the  PPD correction (Figure~\ref{fig:GIS_Pr_before_after}b).
\end{itemize}
\vspace*{-3mm}
\end{table}

\subsubsection{Pulse profiles with the GIS }
\label{subsubsec:GIS_profiles}

In Figure~\ref{fig:GIS_Pr_before_after},
we compare energy-sorted  pulse profiles obtained with the GIS,
before (panel a) and after (panel b) the PPD correction.
They are all corrected for the orbit
using the parameters  in Table~\ref{tbl:orbital_parameters},
and folded at  $P_{\rm fin}$ of Equation~(\ref{eq:P_final}).
In Figure~\ref{fig:GIS_Pr_before_after}(a),
the characteristic three-peak structure is reconfirmed
in the three narrow-band profiles.
At the same time, we notice a clear ``phase-lag" property which
is reminiscent of  Figure~\ref{fig:NuSTAR_PulseProfiles_raw}(a),
except that the phase shift has the opposite sign.
As a result, the broadband (2.8--10 keV)
profile becomes rather dull.
As  in Figure~\ref{fig:GIS_Pr_before_after}(b),
the PPD correction brought  the  narrow-band profiles 
into an excellent mutual phase alignment,
and made them similar to one another,
all clearly exhibiting  the three-peak structure.
The only exception is the 1--2.8 keV result,
where the pulses were not confirmed (\S~\ref{subsubsec:step3})
even when applying the orbit plus PPD corrections. 
(The purple profile for this softest band is shown
without any PPD correction.)}

\begin{figure}
\begin{center}
\includegraphics[width=8.5cm]{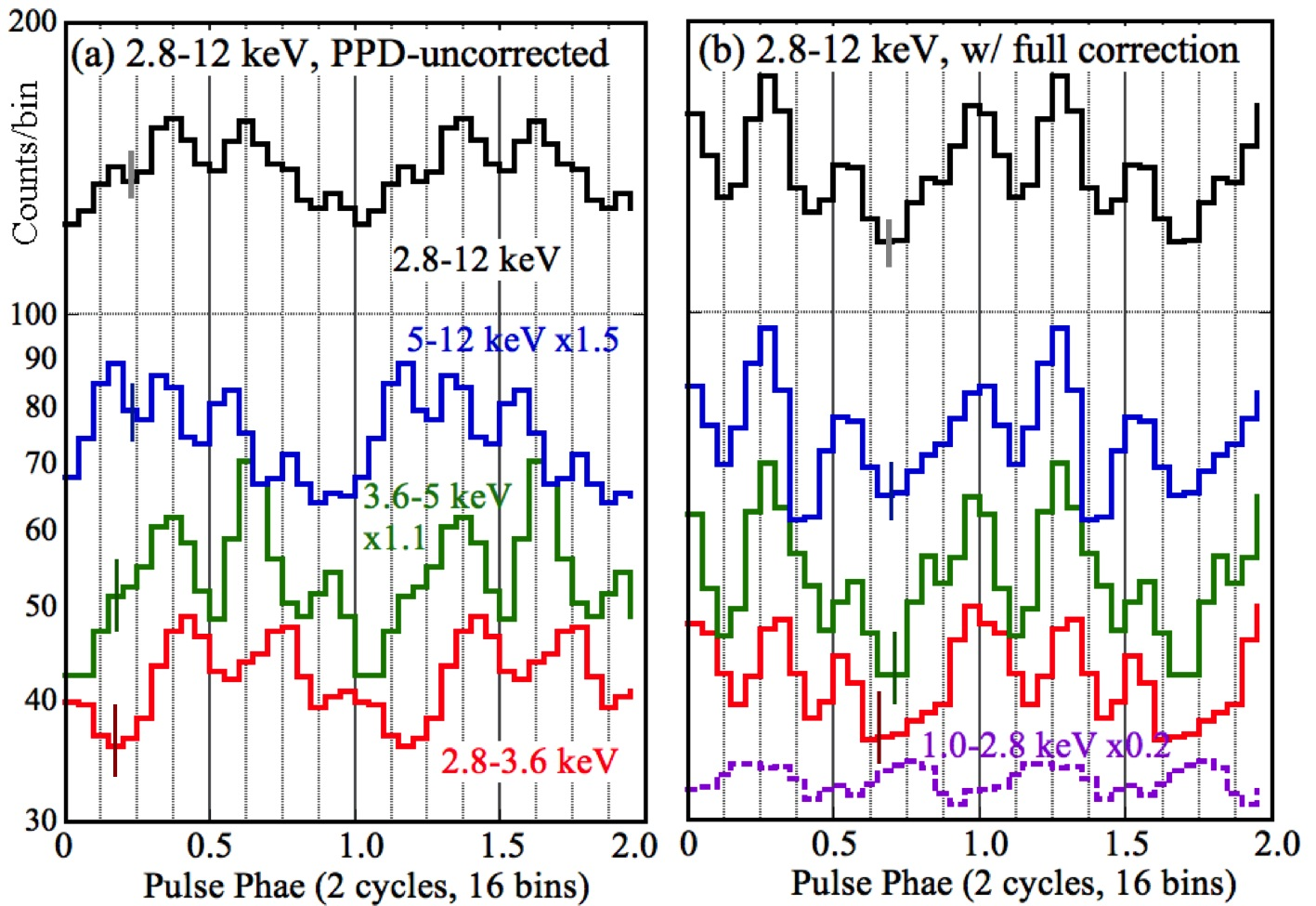}
\includegraphics[width=7.8cm, height=5.2cm]{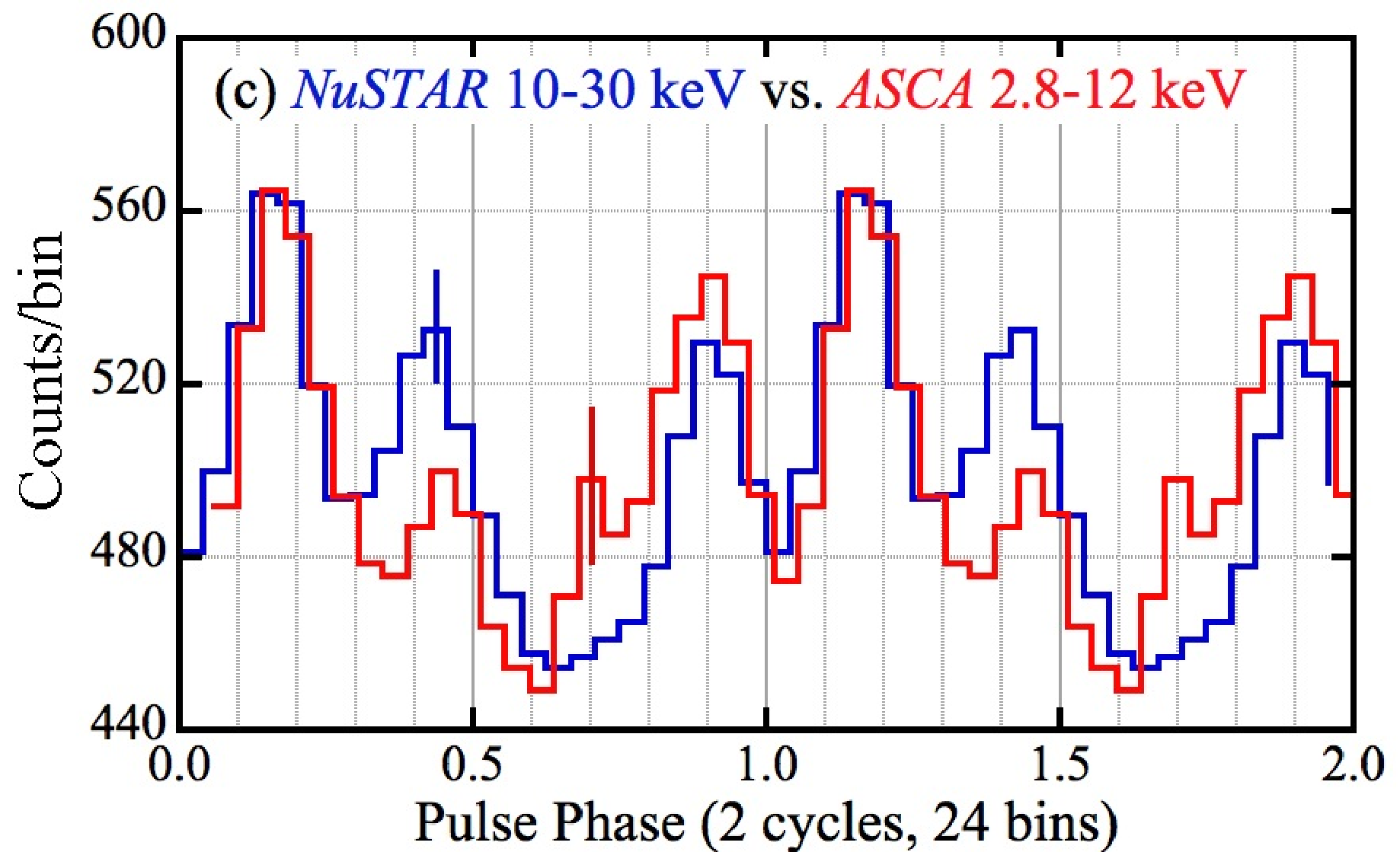}
\end{center}
\caption{
(a) Pulse profiles of \LS\ with the \ASCA\ GIS,
in 2.8--3.6 keV (red),
3.6--5.0 keV (green), 5--12 keV (blue), 
and  2.8--12 keV (black).
They are all orbit-corrected, and folded at $P_{\rm fin}$.
(b) The same, but  further corrected for the  PPD effect,
except in 1.0--2.8 keV (dashed purple).
(c) A comparison between  the GIS (2.8--12 keV; red) 
and \NuS\ (10--30 keV; blue) pulse profiles.
Except the binning,
they are identical to the black histograms in Figure~\ref{fig:GIS_Pr_before_after}(b)
and Figure~\ref{fig:Paper_I_review}(d), respectively.
The GIS profile is rescaled (see text),
and shown with finer bins than in (b).
}
\label{fig:GIS_Pr_before_after}
\end{figure}

Table~\ref{tbl:GIS_profile_eval} summarizes
how the 2.8--12 keV  pulse fraction
increases with the timing corrections.
The  values of 
$\zz$, $\Pch$, and $\chisq$ are also given.
Thus, the combined application of the orbital and PPD corrections
enhances the pulse fraction,
and is essential in detecting the pulsation
in  energies below 10 keV.
However, the increase in the pulse {\it significance} is not obvious here,
because the pre-trial probability $\Pch^{(1)}$
does not consider the trial number 
which evidently increases toward Case 3.

Interestingly,
the final broad-band profile (in black)
and  the 10--30 keV profile with \NuS\
(Figure~\ref{fig:Paper_I_review}a) look very much alike.
In Figure~\ref{fig:GIS_Pr_before_after}(c)
we superpose them together,
after appropriately scaling the GIS data (see below).
The striking agreement between them strengthens
that the GIS and \NuS\ data represent the same phenomenon,
i.e., the source pulsation.

In Figure~\ref{fig:GIS_Pr_before_after}(c),
the pulse-phase origin of the GIS data was shifted by $+0.15$.
This is  because the relative pulse phase cannot be determined 
uniquely between the two data sets separated by 16.9 yr s.
We also rescaled the GIS counts at the $j$-th bin, $C_j$, into  $C'_j$, as
\begin{equation}
C'_j=4.27\left\{ \bar C + 0.54(C_j-\bar C) \right\}
\label{eq:NS_GIS_scaling}
\end{equation}
where $\bar C$ is the average, and the factor 4.27 is the ratio between the
\NuS\ and GIS averages.
The coefficient 0.54, representing ``AC component'',
means that the 10--30 keV  pulse fraction with the \NuS\ is
about half that of the 2.8--12 keV GIS data.
Indeed, the 2.8--12 keV GIS pulse fraction with the full timing corrections, 
$0.21$ (Table~\ref{tbl:GIS_profile_eval}), is 1.6 times
that of the 10--30  \NuS\ profile ($0.135 \pm 0.043$),
although it is much lower than the 10--30 keV \Su\ result of $0.68\pm 0.14$.
This pertains to the issue (iv) raised in \S~\ref{sec:intro},
and suggests that the pulse fraction of \LS\ can vary significantly,
in spite of the stable orbital X-ray light curves.

The profiles  in  Figure~\ref{fig:GIS_Pr_before_after}(b)
use the $e=0.306$ solution in Table~\ref{tbl:orbital_parameters}.
However, even when using the  $e=0.278$  solution,
the profile changes little,
rather than becoming similar to the 10--30 keV HXD profile
 in Figure~\ref{fig:Paper_I_review}(c).
When the SOL-2 parameters 
(Figure~\ref{fig:GIS_OrbCorr}b) are employed,
the three peaks still persist,
but the profile changes considerably,
with the left sub peak becoming highest.

\medskip
\section{Discussion}
\label{sec:discussion}

\subsection{Summary of the data analysis}
\label{subsec:summary}

\subsubsection{Results from the \NuS\ data}

In the present work, we first reanalyzed the \NuS\ data,
focusing on  energies  below 10 keV.
We found that the 9.0538 s pulses lurk
in the data even below 10 keV,
but their detection is hampered by a systematic pulse-phase drift
from $\approx 10$ keV down to at least 3 keV.
Applying  an energy-dependent correction 
to the timings of $\lesssim 10$ keV photons
for this PPD (Pulse-Phase Drift) effect,
the pulses were recovered both in 3--9 keV
and in the 3--30 keV broadband,
where the pulse detection was difficult before the correction.
The PPD effect  in the \NuS\ data is considered real,
because it has  $\Pch  =0.004\%$
(\S~\ref{subsec:NuS_PPD_significance}, Table~\ref{tbl:significance})
in 3--30 keV.

The orbital solution found with the PPD-corrected signals,
either in 3--9 keV or 3--30 keV,
agrees with that from  the hard-band (10--30 keV) \NuS\ data
which are free from the PPD disturbance.
Therefore, the X-ray pulses at $P_0$
are considered to originate from \LS,
both above and below 10 keV.

\subsubsection{Results from the \ASCA\ GIS data}
\begin{figure}
\centerline{
\includegraphics[width=9.5cm, height=5.5cm]{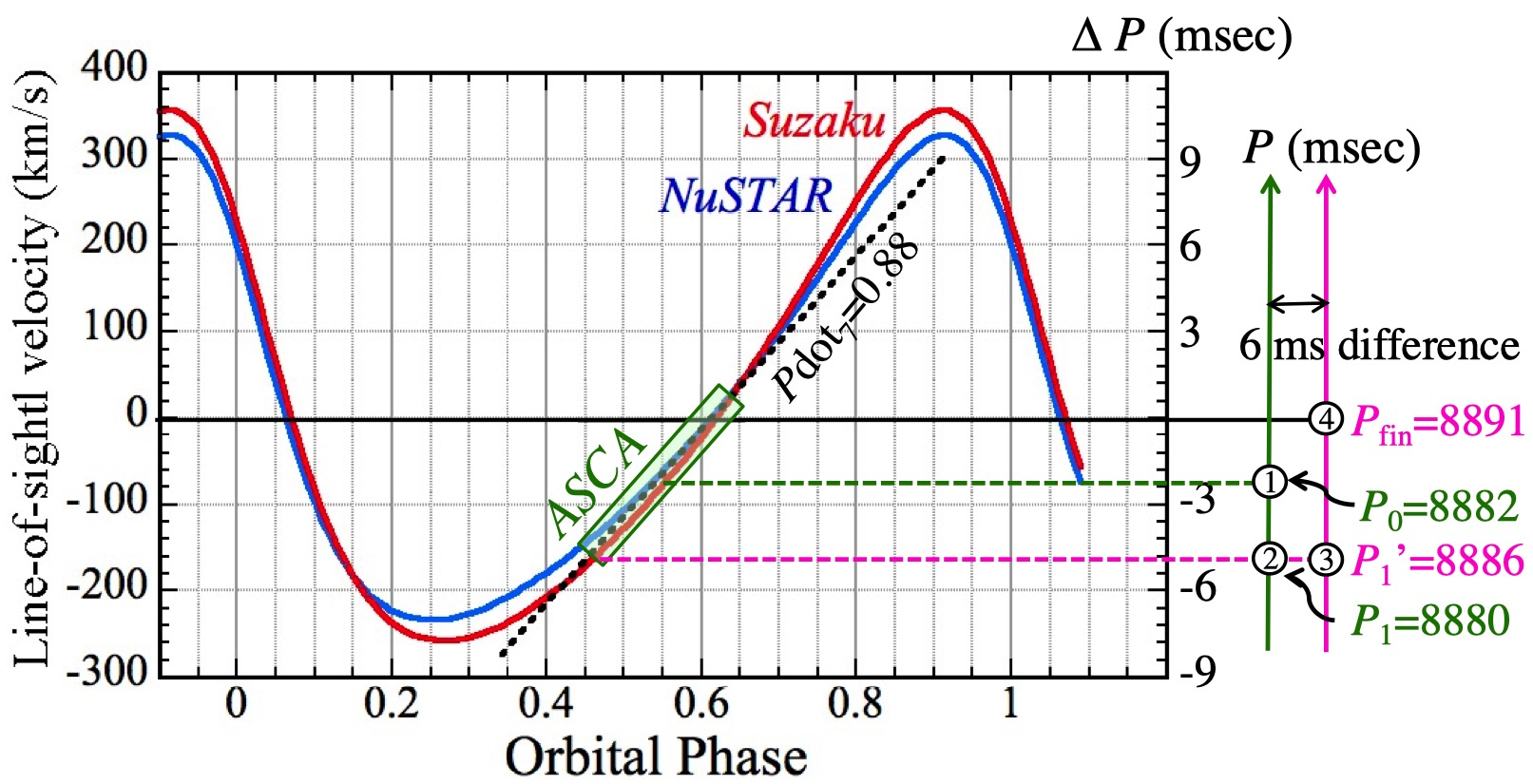}
}
\caption{A summary of the period estimates with \ASCA,
superposed on the radial X-ray Doppler curve
from the \Su\ (red) and \NuS\ (blue) solutions.
Periods and period differences are all shown in units of ms.
The periods found in Stage 1 through Stage 4 are indicated
by circled numbers. 
The orbital phase covered with \ASCA\ is indicated in green,
and the optimum $\dot P_7$ by a dotted black line.
}
\label{fig:Doppler_curve}
\end{figure}

We next analyzed the \ASCA\ GIS data of \LS\ in four Stages,
as depicted in Figure~\ref{fig:Doppler_curve}
superposed on the predicted orbital Doppler curves.
In Stage 1 (\S~\ref{subsubsec:step1}),  
a simple $m=2$ PG  in 6--12 keV,
covering a wide period range of $P=8.2-9.2$ s, 
gave evidence for periodicity at $P_0$ of Equation~(\ref{eq:P0})
(dashed green line in Figure~\ref{fig:Doppler_curve}).
It  has  $\Pch \approx 16\%$ or $\approx 1.2\%$
(Table~\ref{tbl:significance}),
when evaluated in the  period interval of 8.2--9.2 s,
or 8.82--8.90 s (consistent with
Equation~\ref{eq:ASCA_P_prediction}), respectively.

The Stage 2 analysis (\S~\ref{subsubsec:step2}), 
using the 5.3--12 keV photons, 
$m=4$, and the 8.82--8.90 s range,
mimicked the orbital Doppler curve by a constant  $\Dot P$.
Then, $\zz$ increased from 25.1 (Stage 1) to 32.1,
and the  period  was revised from $P_0$ 
to $P_1$ of Equation~(\ref{eq:GIS_two_solutions}).
The latter is shown by a magenta line,
and its shift  from $P_0$ agrees with the expected Doppler velocity difference
between the mid point and the start of the observation.

In Stage 3 (\S~\ref{subsubsec:step3}),
we identified  the PPD effect below 10 keV
with high significance ($\Pch \approx 0.2\%$; 
Table~\ref{tbl:significance}, Appendix C), 
and performed its correction.
The periodicity then became significant in the 2.8--12 keV broadband,
when  $\dot P_7 \approx 0.88$ (dashed oblique  line) is employed.
The period changed from $P_1$  to $P_1'$ of Equation~(\ref{eq:Pdot2}a),
by  6 ms which can be explained by interference with $2 P_{\it sc}$.
In Stage 2 and Stage 3, the approximation of the orbital effects 
by a constant $\dot P$ worked  successfully. 

The final (4th) step was to apply the full orbital correction to the 2.8--12 keV data,
together with  the PPD correction found in Stage 3.
Then, the period found in Stage 1 was finally refined as  $P_{\rm fin}$ of Equation~(\ref{eq:P_final}),
which falls on the zero-velocity abscissa in Figure~\ref{fig:Doppler_curve}.
The folded profile became much more structured 
than  with partial  or no timing  corrections (Table~\ref{tbl:GIS_profile_eval}).
This $P_{\rm fin}$,
lying approximately on the \NuS\ to \Su\ back-extrapolation
(\S~\ref{subsec:period_history}), can be identified 
readily with the  pulsation of \LS\ found in Paper I.
This  identification is reinforced by  
nearly the same values of $\Eb$ between \NuS\ and \ASCA, 
the striking  similarity between the \ASCA\ and \NuS\ pulse profiles 
(Figure~\ref{fig:GIS_Pr_before_after}c),
and the consistency (Table~\ref{tbl:orbital_parameters}) 
of the \ASCA\ specified orbital parameters
with those from \NuS\  (if assuming $e=0.306$)
or \Su\  (if assuming $e=0.278$).
We thus conclude that the \ASCA\ data provide another  evidence
of the pulsation from \LS,
although the issues (iii) and (iv) remain unsolved.

\subsection{PPD effects}
\label{subsec:PPD}

Evidently, the key concept in the present work is 
the PPD phenomenon, 
noticed both in the \NuS\ and \ASCA\ data below 10 keV.
Though very enigmatic, 
its reality is support by the following arguments.

\begin{enumerate}
\item
Such an effect, though rather rare,  was 
observed with \NuS\ by \cite{Miyasaka13}, 
from the Be binary  GS 0834$-$430 during its 2012 outburst.
Similar behavior was also detected from two magnetars,
\oneE\ \citep{Makishima21a} and SRG~1900+14 \citep{Makishima21b}.
\item
As in  Figure~\ref{fig:NuSTAR_PulseProfiles_raw}
and Figure~\ref{fig:GIS_Pr_before_after}(a),
the phenomenon in \LS\ was first  revealed by the data themselves, 
in a data-oriented analysis  
that is not biased by any  particular preconceptions.
\item
It is confirmed with high statistical significance
in both the \NuS\ and \ASCA\ data  (Appendix B, Appendix C),
at a consistent break energy of $\Eb \approx 10$ keV.
\item
Although the difference  between
$\R \approx 60$ with \NuS\ and  $\R \approx -24$ with \ASCA\ is puzzling,
a similar sign change was seen in SGR 1900$-$14 \citep{Makishima21a}.
\end{enumerate}

We  thus successfully ascribed the 
soft X-ray pulse suppression to the PPD perturbation.
Therefore, a part of the doubts on Paper I \citep{Volkov21} was removed,
and the issue (ii) in \S~\ref{sec:intro}  was  {\it half} solved.
The remaining half is to identify the  astrophysical origin 
of this phenomenon.
Some attempts are carried out in \S~\ref{subsec:PPD_astro},
though still inconclusive.

\subsection{Significance of the 9 s pulsation}
\label{subsec:Pns_reinforce}

Referring to Table~\ref{tbl:significance},
the issue (i) is considered.
We take the pulsation in the  \Su\ data  for granted,
because it has $\Pch\sim 0.15\%$
against a wide period-search range of 1--100 s,
and is reconfirmed  independently 
by \citet{Volkov21} using the same data.

Let us first evaluate how the present \NuS\ results reinforce
the pulse significance  in the \NuS\ data.
In 3--9 keV, the PPD significance was estimated to be 
$\Pch \sim 5\%$  (\S~\ref{subsec:NuS_PPD_significance}),
but it can also be regarded as the chance probability
to observe  $\zz>36.22$ at $\Pns$.
Because the 3--9 keV and 10--30 keV results
utilize independent photon sets,
we are allowed to  multiply this  $\Pch$  in 3--9 keV
onto that from  10--30 keV, $\Pch \sim 7\%$,
to revise the estimate as $\Pch \sim 0.35\%$.

Alternatively,  the 3--30 keV result in Table~\ref{tbl:significance} may be utilized,
but it cannot be combined with that in 10--30 keV,
because  the  energy intervals overlap.
Then, we return to  the 9.025--9.065 s period range,
which contains 165 independent Fourier waves
(for the data span of 3.9 d).
Multiplying the 3--30 keV estimate of $\Pch \sim 0.004\%$ by 165,
we obtain $\Pch \sim 0.7\%$,
which is comparable to the first estimate.
In either case, the probability for the \NuS\ pulse detection to be false
diminishes by an order of magnitude from Paper I,
with the pulse confidence increasing to  $>99\%$.

We next consider the \ASCA\ data analysis,
where the significance of the $\sim 8.9$ s pulses 
was evaluated at Stage 1 and Stage 3.
Of them, we choose the Stage 3 estimate, $\Pch\sim 0.2\%$,
because it is based on the largest number of GIS photons
available for the pulse search.
Again, this $\Pch$ primarily means a  probability for 
the PPD effect (\S~\ref{subsubsec:step3}),
but it can also be regarded as the probability
to observe, through the PPD correction,
a value of  $\zz>53.6$ by chance.
It takes into account the trials in $P$ (over 8.82--8.90 s),
$\dot P$ (substituting for $\ax$, $\omega$, and $\tau_0$), 
$\R$, $\Eb$, and $E_{\rm L}$ (Appendix C).

Thus, the chance probability 
of the $\sim 9$ s pulsation in \LS\ is estimated as
$\Pch\sim 0.15\%$ with \Su\ (Paper I),
$\sim0.35\%$ or $\sim 0.7\%$ with \NuS,
and $\sim0.2\%$  with \ASCA.
We refrain, however, from combining these values of $\Pch$,
as they may not be  independent of one another.
Furthermore, deriving too small values of $\Pch$ would be meaningless
when considering  various systematic uncertainties;
e.g., the unsolved issues of (iii) and (iv), the remaining half of (ii), 
possible biases in Equation~(\ref{eq:ASCA_P_prediction}),
and the fact that the initial \Su\ search 
did not cover the period range below 1 s.

In any event,
the $\sim 9$ s pulsation of  \LS\ is thought 
to be significant  with $>99\%$ confidence
in all the three data sets.
This affirmatively settles the objective (i),
and strengthens the conclusion in Paper I,
that the compact object in \LS\ is a magnetized NS.
The  mass estimate of $1.23-2.35$ Solar masses,
derived in Paper I assuming the pulsation to be real,
is consistent with this conclusion.

\subsection{Long-term spin down of \LS}
\label{subsec:period_history}
\begin{figure}
\centerline{
\includegraphics[width=6.5cm]{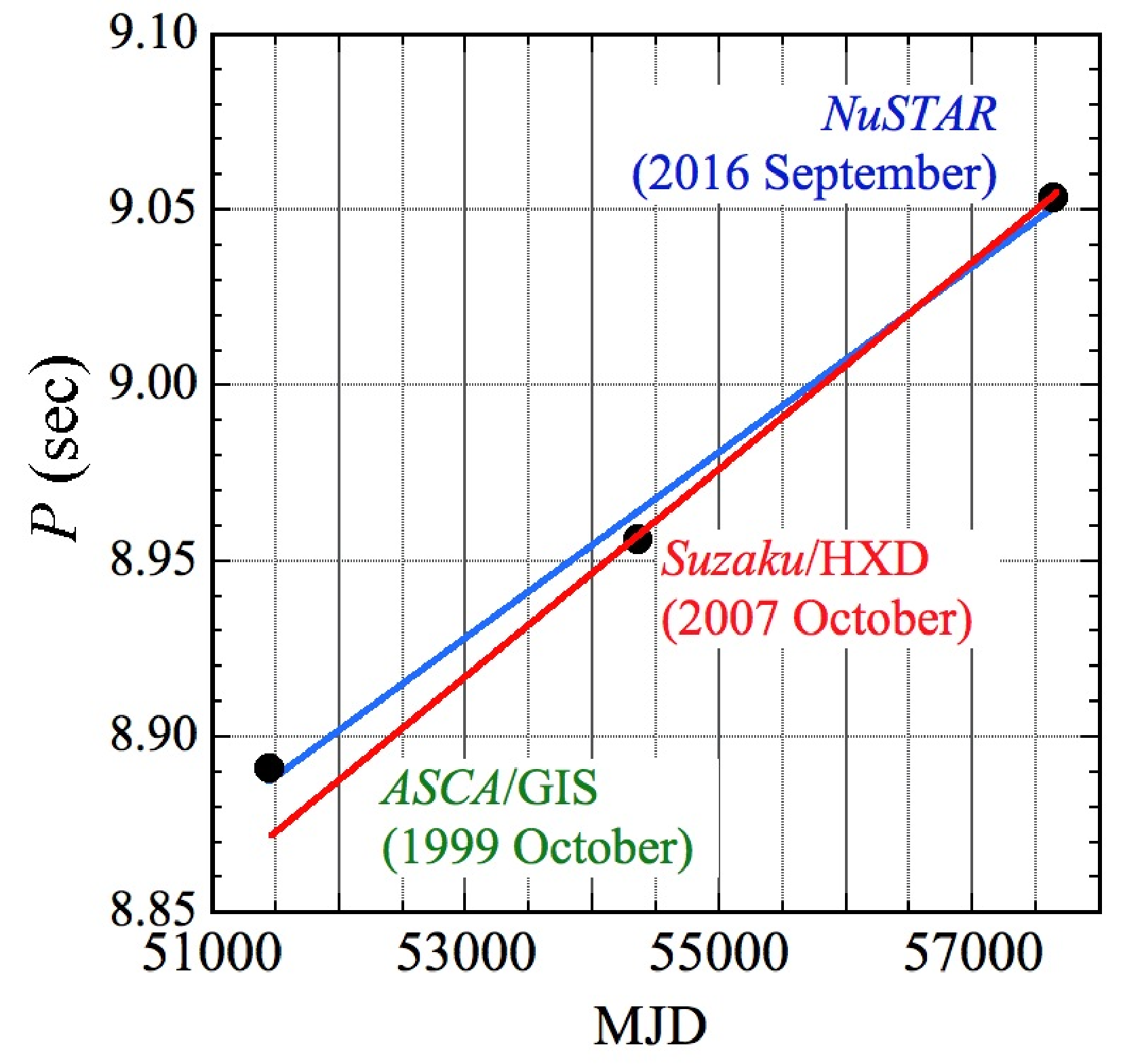}
}
\caption{The pulse period of  \LS\ measured with \ASCA,
compared with those from \Su\ and \NuS.
The red line connects the \Su\ and \NuS\ data points,
where the blue one represent  an average spin-down trend.
}
\label{fig:period_history}
\end{figure}

Taking the pulsation for granted,
the pulse periods of \LS, measured for 17 yrs from 1999 to 2016 with \ASCA, \Su, and \NuS,
are plotted  in Figure~\ref{fig:period_history}.
We  derive $\dot P= 2.6 \times 10^{-10}$ s s$^{-1}$ from  \ASCA\  to \Su, 
and  $\dot P= 3.4 \times 10^{-10}$  s s$^{-1}$ from \Su\ to \NuS.
Thus, $\dot P$ is inferred to change mildly,
around an average rate of  $3.0 \times 10^{-10}$ s s$^{-1}$
which  implies a characteristic age of about 480 yr.
The system is indicted to be very young, 
in agreement with the fact that the optical companion is a massive star.

Now that not only $P$ but also $\dot P$ reported in Paper I were reconfirmed,
major discussions and conclusions given there remain intact.
Namely, the  bolometric luminosity $\sim 10^{36}$ erg s$^{-1}$ of the source
can be powered by neither  the rotational energy loss nor mass accretion.
The stellar winds cannot provide the required energy input, either.
Therefore, the compact object in \LS\ must be
a magnetically powered NS, or a magnetar in a binary.
The lack of mass accretion from the stellar winds  
can be explained by assuming that the Alfv\'en radius  exceeds
the gravitational  wind-capture radius.

The observed $\dot P$  could  be 
explained by emission of magnetic dipole radiation,
if the dipole magnetic  field reaches $B \sim 10^{15}$ G  (Paper I).
However, this  cannot be the dominant spin-down mechanism,
because it would predict  $\dot P$ to decrease with time,
in disagreement with the present result.
As a more likely scenario,
the object may spin down via  interactions with the stellar winds (Paper I),
and fluctuations in this process produce the mild variation in  $\dot P$.
At the same time, these interactions presumably cause
the magnetar's  magnetic energy to somehow dissipate 
into particle acceleration and the gamma-ray emission \citep{Yoneda21}.
Admittedly, however, this scenario is still subject to many unknowns,
including where the X-ray emission in fact comes from,
how the NS's magnetic energy is converted to that of the particle acceleration,
and how this process is ``catalyzed" by the stellar winds.

\subsection{The nature of the PPD phenomenon}
\label{subsec:PPD_astro}

Let us consider possible astrophysical origins of the PPD phenomenon,
after \cite{Miyasaka13} who discussed  various possibilities 
to explain the behavior of GS 0834$-$430.
Except this Be pulsar, \LS,  and  the two magnetars mentioned in \S~\ref{subsec:PPD},
this phenomenon  is rather rare among compact X-ray sources.
Therefore, it may take place under some limited  conditions,
e.g., the presence of magnetar-class magnetic fields
which \LS\ is suggested to harbor (Paper I).

The PPD observed from \LS\ has three charcteristics;
(1) $\Eb \approx 10$ keV agrees  between the two data sets;
(2) between them, the sign of $\R$ is opposite;
(3) the effect suddenly sets in at $E = \Eb$,
below which the pulse phase depends linearly on $E$.
Of these, (2) rules out
reprocessing of harder photons into softer ones 
after some delays (soft lag),
or hardening of softer photons by, e.g., 
Compton up-scattering  (hard lag).
In contrast, as proposed by \cite{Miyasaka13},
an energy-dependent beaming of the X-ray emission may work.
Then, (1) and (3) suggest 
that the threshold energy $\Eb$ has a certain physical meaning,
across which the properties of  photon emission and/or transfer
change distinctly.

One possibility to explain $\Eb$ is electron cyclotron resonance \citep[e.g.,][]{Harding99}
in the  strong magnetic fields somewhere around the NS.
However, as discussed in Paper I in the magnetar context,
the stellar winds flowing toward the NS would be interrupted
at the Alfv\'en radius, $R_{\rm A} \sim 2 \times 10^{10} B_{15}^{1/3}$ cm,
where $B_{15}$ is the surface magnetic field of 
the assumed magnetar in units of $10^{15}$ G.
At $R_{\rm A}$, the magnetic field would decrease to
$B_{\rm A} \sim B_{15}(R_{\rm NS}/R_{\rm A})^3 
\sim 1.3 \times 10^7$ G regardless of $B_{15}$,
where $R_{\rm NS} \sim 10^{6}$ cm is a typical NS radius.
This $B_{\rm A}$ is five orders of magnitude lower than 
would explain the electron cyclotron resonance at $\sim 10$ keV.

Putting aside (3),
a mechanism to deflect the X-ray propagation direction,
in an energy-dependent way, is well known in laboratory;
namely, X-ray diffraction techniques.
Taking a transmission grating as an example,
the 1st order diffraction beams are deflected,
from the indecent beam direction,
by an angle which is inversely proportional to the X-ray energy.
This angle has opposite signs between 
beams of the order $+1$ and $-1$,
as required by (2).
However, in the present astrophysics setting,
it is totally unclear what serves as the diffraction grating.
Furthermore, the energy dependence in this scenario
differs from what is seen in \LS.

In this way, satisfactory explanations 
of the PPD phenomenon are yet to be sought for.
Nevertheless,  it may possibly give a clue
to strong-field physical phenomena 
that may be taking place in this interesting object.
Future X-ray polarimetric observations are encouraged.

\subsection{The \Su\ versus \NuS\ discrepancy}
Although the issues (i) and (ii) have been solved at least partially,
we are still left with (iii) and (iv),
or in a word, \Su\ {\it vs.} \NuS\ discrepancy.
As to (iv),  at present we can say
only that the pulse fraction of this source may vary considerably
for unknown reasons.

Let us consider (iii). It is not due to underestimations 
of errors (Paper I) associated with the orbital parameters.
In fact, when $e$, $\ax$, and $\omega$ of the  \Su\ solution
are forced onto the 10--30 \NuS\ data, the peak $\zz$, which was 66.87
(Table~\ref{tbl:orbital_parameters}), worsens to 29.6, 
even when allowing ample tolerances for $P$ and $\tau_0$.
Vice versa, the peak $\zz$ of the \Su\ data degrades from 67.97 to 26.4, 
when $e$, $\ax$, and $\omega$ are fixed to the \NuS\ solution
while $P$ and $\tau_0$ are allowed to vary widely.
The discrepancy is statistically significant.

The present  \ASCA\  results provide a possible clue to this issue.
The maximum value of $Z_4^2=50.7$ or 49.9 (Table~\ref{tbl:orbital_parameters}),
achieved in Stage 4, is puzzlingly lower than that ($Z_4^2=53.6$)
obtained in Stage 3 where the orbital Doppler changes in $P$
are approximated by a constant $\dot P$.
Obviously, this is opposite to the  expectation,
that the full orbital correction in Stage 4 should be more accurate,
and would give a higher $Z_4^2$.
Then, the actual Doppler curve might be slightly deviated 
from those predicted by an ideal elliptical orbit,
and happened to be more straight during the \ASCA\ observation.
Such perturbations can arise if, e.g., \LS\ is a triple system 
with an unseen third body \citep[e.g.,][]{3body_system},
or the X-ray emission region is somewhat ($\sim $ a few lt-s) displaced from the NS,
or the one-day period variation in the radial velocity of 
the optical companion \citep{Casares11} contributes,
or the NS undergoes free precession.

Among the above possibilities, we favor the free-precession scenario,
because nearly all magnetars are deformed
by their internal magnetic pressure to an asphericity of $\epsilon \sim 10^{-4}$,
and hence its rotation and precession periods become different 
by $\approx \epsilon$ in fraction (e.g., \citealt{IAU363}).
The two periods interfere with each other,  to modulate the pulse phase
at their beat period, $T_{\rm beat} \approx P/\epsilon$.
If $\epsilon \sim 0.3 \times 10^{-4}$, we find $T_{\rm beat} \sim 3.4$ d,
which is comparable to the orbital period (3.906 d) of \LS.
When this effect is superposed on the orbital modulation,
the Doppler curve will vary to some extent from epoch to epoch,
and could explain the \Su\ and \NuS\ discrepancies.

\section{Conclusions}
By revisiting the \NuS\ data of \LS\ after Paper I,
and analyzing the \ASCA\ data taken in 1999,
we have arrived at the following conclusions.

\begin{itemize}
\item
In the \NuS\ data at energies below 10 keV, 
a PPD (Pulse-Phase Drift) effect is noticed with high significance.
Its correction recovered the pulsation in 3--9 keV and  3--30 keV,
and strengthened the pulse detection with \NuS,
although astrophysical origin of this  phenomenon is unclear.
\item
When the PPD and orbital corrections are  incorporated,
the  2.8--12 s  \ASCA\  GIS data acquired in 1999
gave evidence for a  8.891 s pulsation. This result,
when combined with those from \Su\ and \NuS,
further reinforces the reality 
of the $\sim 9$ s pulsation in \LS.
\item
For 17 yrs from 1999 to 2016,
the object has been spinning down with an average rate 
of $\dot P = 3.0 \times 10^{-10}$ s s$^{-1}$,
although $\dot P$ may be mildly variable.
The characteristic age becomes 480 yr.
\item 
Through the new pulse-period measurement
and reconfirmation of $\dot P$,
the present work validates all discussions in Paper I
and \cite{Yoneda21},
that the compact object in \LS\ is a magnetized NS 
with a pulse period of $\sim 9$ s, 
and is likely to be magnetically powering the non-thermal radiation.
\item
The pulse-period change along the binary phase could be subject to
some perturbations besides the simple Doppler effect
in an elliptical orbit.
\end{itemize}

\medskip
\section*{Appendix A: The chi-square and $Z^2$ statistics}

Both   $Z_m^2$  \citep{Z2_94} and $\chisq$ quantify 
the deviation of a pulse profile from a flat one.
While $\chisq$ utilizes  profiles folded into $N$ bins,
$Z_m^2$ operates on discrete photon data,
by adding the power from fundamental 
to the specified maximum harmonic, $m$.
If  $m=N/2$, the two statistics become the same  due to the Parseval's theorem.
For white noise data, $Z_m^2$ obeys a chi-square distribution
with $\nu=2m$ degrees of freedom.

\begin{figure}
\centerline{
\includegraphics[width=8.9cm]{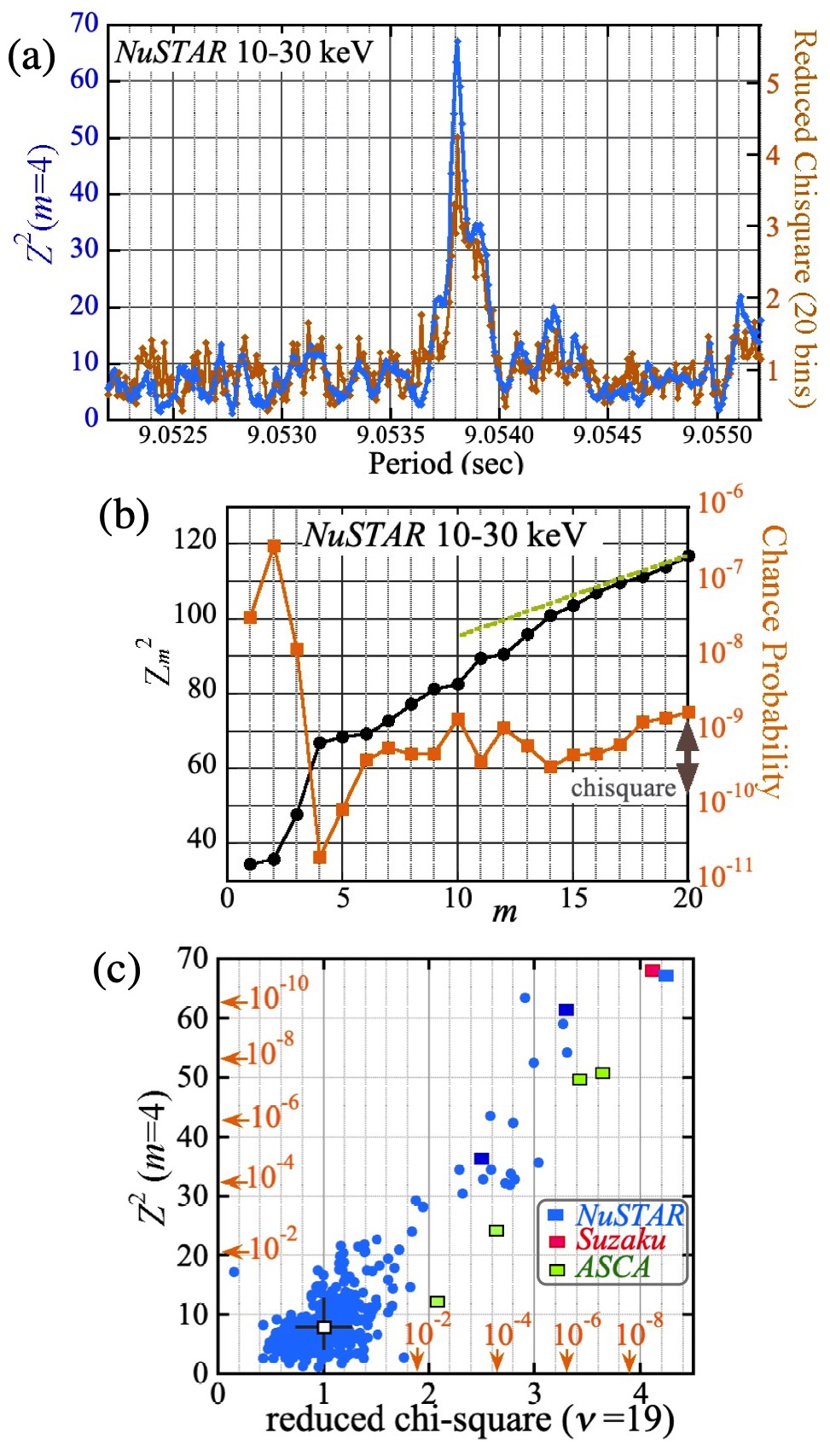}
}
\caption{
Comparison of the $\chisq$ and $Z_m^2$ statistics.
(a) The orbit-corrected \NuS\ PGs in 10--30 keV,
calculated by $Z_4^2$ (blue) and $\chi_{\rm r}$  (brown; $\nu=19$).
(b) The maximum $Z_m^2$ (black) from the 10--30 keV \NuS\ data,
shown as a function of $m$. 
The associated $\Pch^{(1)}$ is given in orange, 
where the vertical brown arrow indicates typical values obtained with $\chisq$.
(c) A scatter plot between $\chi_{\rm r}$ and $Z_4^2$,
for the results in panel (a), Table~\ref{tbl:orbital_parameters},
and Table~\ref{tbl:GIS_profile_eval}.
The white box with black  bars indicates
the average and standard deviation for white noise.
The values of $\Pch^{(1)}$ are shown in orange.
}
\label{fig:Chisq}
\end{figure}

Figure~\ref{fig:Chisq}(a) compares two PGs
from the 10--30 keV \NuS\ data with the orbital correction.
The blue one,  using $\zz$, is similar to Figure~\ref{fig:Paper_I_review}(b),  
but the peak is  sharper,
because we fix the orbital parameters (Table~\ref{tbl:orbital_parameters}).
The other PG in brown uses $\chisq$ with $N=20$ ($\nu=19$).
The two ordinates are rescaled
so that their averages and standard deviations 
become about the same.
Although the two PGs look alike,
including the peak at $\Pns$, 
we can point out three advantages of $Z_m^2$ over $\chisq$.

First, the peak value of $\chisq=4.24$ in Figure~\ref{fig:Chisq}(a)
gives a single-trial probability $\Pch^{(1)}=1.5 \times 10^{-9}$,
which  is much worse than that ($\Pch^{(1)} = 2.1 \times 10^{-11}$) from $\zz =66.87$.
This is because the power in the pulses of typical X-ray pulsars 
(except, e.g,  the Crab pulsar with very sharp profiles)
is limited to the first several harmonics,
beyond which the noise power dominates.
Hence, $\chisq$ is affected by these noise contributions
summed up to the Nyquist harmonic.
Actually in Figure~\ref{fig:Chisq}(b),
$Z_m^2$ for the \NuS\ data is seen to increases rapidly until $m=4$,
above which it approaches a prediction for 
white noise data (dashed green line).
As a result, $\Pch^{(1)}$ from $Z_m^2$  takes a clear minimum at $m=4$.
Figure~\ref{fig:Paper_I_review}(c)
compares $\zz$ and $\chisq$ from the present analysis (see caption).
Again, $\chisq$  gives a systematically higher $\Pch^{(1)}$  than $\zz$,
even though a roughly linear relation holds between the two quantities.

The second point is that $\chisq$ and the associated $\Pch^{(1)}$
both depend considerably on $N$,
and there is no clear principle as to which $N$ to choose.
In fact, the peak in Figure~\ref{fig:Chisq}(a) has
$\chisq=3.62$, 4.24, and 5.06, for $N=24$, 20, and 16, 
implying $\Pch^{(1)}/10^{-9}=9.3, 1.5$, and 0.39, respectively.
In contrast, $Z_m^2$ is free from this problem,
because it is based on unbinned likelihood evaluation,
and does not need the data binning;
it  is hence suited to photon time-series data.
In practice, we fold the data into profiles of $N=120$ bins,
but this is a convention to accelerate the $Z_{\rm m}^2$ computation,
and the derived $Z_{\rm m}^2$ does not depend on $N$ as long as $m \ll N$.

Finally in Figure~\ref{fig:Chisq}(a),
the $\chisq$ PG has a shorter coherence length than 
that of $\zz$, with a sharper pulse peak.
Evidently, this is because $\chisq$ takes into account
the smallest fine structures 
(mostly due to noise) in the profile.
When calculating a PG with $\chisq$, 
we hence need to employ much finer parameter grids,
so as not to miss the peaks.
As a result, the computational times 
can easily increase by an order of magnitude or more.

For these reasons, in either principle or practice,
we keep using the $Z_m^2$ statistics.
In any case, the evaluation of pulse significance
should always refer to $\Pch^{(1)}$,
rather than the face values of $\chisq$ or $Z_m^2$.

\section*{Appendix B: Significance of  the PPD effect in the NuSTAR data}

Based on the prospect in \S~\ref{subsec:NuS_PPD_significance},
we evaluated the statistical significance 
of the PPD effect in the \NuS\ data below 10 keV,
via a {\it control} study using the data themselves.
Namely, we calculated 3--9 keV PGs 
such as Figure~\ref{fig:NuSTAR_Pscan_PPD}(a),
over two  period ranges,
$P=8.0-8.5$ s and $P=10.2-10.7$ s, avoiding a vicinity of $\Pns$.
To ensure that adjacent period samplings are Fourier independent,
$P$  was changed with a step of 1 ms,
to achieve 1000 period samplings.
At each $P$, we  readjusted $\tau_0$, $\ax$, and $\omega$
as in Figure~\ref{fig:NuSTAR_Pscan_PPD}(a),
and further  allowed $\R$ to vary from $-90$ to $+90$, 
with a step of 1.0 (all in deg keV$^{-1}$).
As in \S~\ref{subsec:NuS_PPD_significance},
$\Eb=10$ keV was fixed.
Out of the 1000 samplings,
in 48 cases $\zz$ exceeded 36.22, the target value.
Thus, the  peak in  Figure~\ref{fig:NuSTAR_Pscan_PPD}(b)
will appear by chance with a probably of $\Pch \sim 5\%$.
Here, the  evaluation at a single period is adequate,
because we consider only the periodicity at $P_0$.

\begin{figure}
\centerline{
\includegraphics[width=8.3cm]{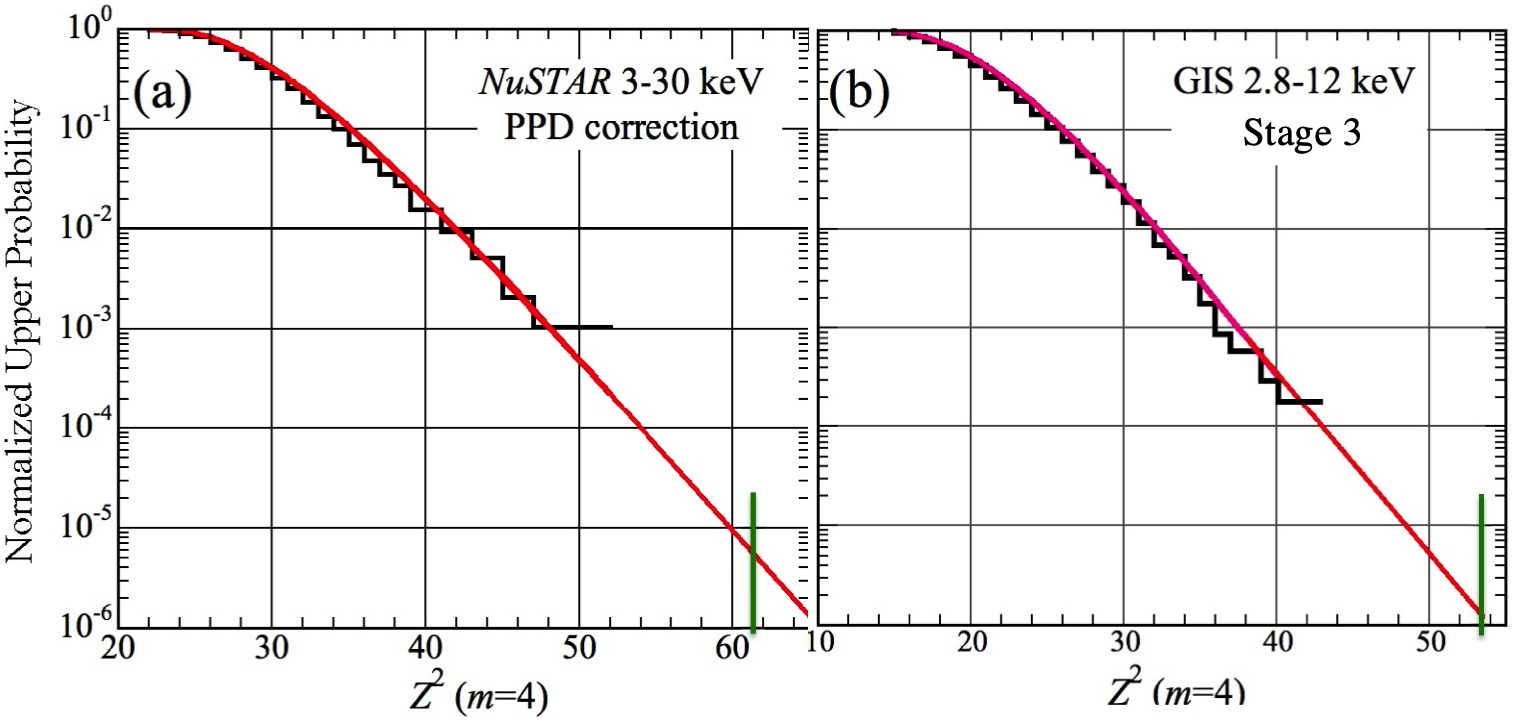}
}
\caption{Upper probability distributions of $Z_4^2$ values
from the {\it control} studies using off-peak period ranges.
At each period, the PPD correction is incorporated,
with $\R$ allowed to vary freely
in the same manner as for the pulse-period search.
See text (Appendix C) for details.
The target values are indicated by vertical green line segments.
(a) For the 3--30 keV \NuS\ data
incorporating the orbital and PPD corrections,
with $\Eb=10.4$ fixed.
(b) For the 2.8--12 keV GIS data in Stage 3,
assuming  $\Eb=10$ keV.
}
\label{fig:UpperProb}
\end{figure}

The evaluation was performed also in the 3--30 keV broadband,
using the same procedure,
again to achieve 1000 period samplings in total.
Of them,  the maximum was 52.36,
which is lower than the target value, 61.51.
To accurately extrapolate the constraint,
we converted these samplings into a distribution 
of the integrated upper probability ${\cal P} (>Z_4^2)$
\citep{Makishima14,Makishima16},
i.e., a probability that $Z_4^2$ in  a single trial 
exceeds that value due to chance fluctuations.
The result is given in  Figure~\ref{fig:UpperProb}(a),
where the  last data bin is at ${\cal P} (>52.36)=1/1000$.

We fit the distribution with an empirical function
\begin{equation}
y = \exp \left[ a - d^{-1} \sqrt{(x-b)^2+c^2}\right]
\label{eq:UpperProb}
\end{equation}
where $x$ and $y$ stand for $\zz$ and ${\cal P}$,
respectively, while $a, b, c$, and $d$ are adjustable parameters.
For $x\gg c$, it  approaches $y \propto \exp(-x/d)$,
which represents a chi-square distribution if $d=2$.
For $|x| \ll c$, it reduces to a shifted Gaussian centered at $x=b$.
As shown in Figure~\ref{fig:UpperProb}(a) by a red curve,
a reasonable fit for $x\geq20$ was obtained 
with $a=4.51$, $b=22.94$, $c=10.85$, and $d=2.4$.
The value of  $d$, somewhat larger than 2.0,
is due probably to some non-Poissonian variations.

When the fit is  extrapolated,
we obtain ${\cal P}(>61.5) = (5.1 \pm 0.5) \times 10^{-6}$.
While it refers to a single value of $\Eb=10.4$,
Figure~\ref{fig:NuSTAR_Pscan_PPD}(c) was derived by 
varying $\Eb$ from $9.0$ keV to $11.0$ keV.
Although we performed 20 steps in $\Eb$,
only $\sim 7$ of them are estimated to be independent,
from the error in $\Eb$.
Then, multiplying ${\cal P}(>61.5)$ by 7,
we obtain a probability of $\Pch \sim 3.6 \times 10^{-5}$, or 0.004\%,
for a value of $\zz \geq 61.5$ to appear by chance at $P=\Pns$.
Since this  $\Pch$ is sufficiently small,
the PPD effect is concluded to be statistically significant.

A concern is that we may have  missed peaks of $\zz$, 
due to the sparse samplings in $P$
employed to make them  independent.
We hence repeated the same study
using a finer period step of 0.2 ms.
However, the result was essentially the same;
${\cal P}(>61.5) = (3.8 \pm 0.9) \times 10^{-6}$.
Presumably, the concern is avoided by utilizing 
the whole distribution of  integrated upper probability,
instead of relying upon the highest $\zz$ value.

\section*{Appendix C: Significance of  the GIS Stage 3 periodicity}

Let us  evaluate the statistical significance 
of the \ASCA\ periodicity found in Stage 3 
at $P_1'$ with $Z_4^2=53.6$ (Figure~\ref{fig:GIS_Rscan}b),
incorporating the PPD correction.
The analytic method, 
employed  in Stage 1 to estimate the significance of $P_0$,
is here inapplicable,
because the effective number of trials in $\R$
cannot be easily evaluated.
We hence follow Appendix B, 
and calculated $m=4$ PGs,
utilizing the  same 2.8--12 keV GIS data,
but over two offset period ranges;
7.0--8.0 s (CNTL-1), and 9.0--10.0 s (CNTL-2),
which avoid the period range of Equation~(\ref{eq:ASCA_P_prediction}).
We  scanned $P$ with a step of 1.0 ms for CNTL-1 (1000 samplings),
and 1.4 ms for CNTL-2 (714 samplings).
Like in Figure~\ref{fig:GIS_Rscan}, $E_{\rm b}$ was fixed at 10.0 keV, 
and $\R$ was allowed to vary, at each $P$, 
in the same manner as  in Stage 3.
The same three values of $\dot P$ as in Figure~\ref{fig:GIS_Rscan} were tested; 
$\dot P_7 =0.75,  0.9$, and 1.05.
Then, we assembled all $Z_4^2$ values from CNTL-1 and CNTL-2, 
as well as over the three cases of $\dot P$.
Thus,  in total  $(1000+714) \times 3=5142$ samplings of $Z_4^2$ were obtained.
The largest of them was 43.5,
much smaller than the target value of $53.6$.

As in Appendix B, 
these samplings were rearranged into a distribution of ${\cal P} (>Z_4^2)$
shown in  Figure~\ref{fig:UpperProb}(b).
The last data bin is at ${\cal P} (>43.5)= 2.0 \times 10^{-4} =1/5142$.
The  distribution was fitted again with  Equation~(\ref{eq:UpperProb}),
but with $d=2.0$ fixed.
As superposed in red, a reasonable fit was again obtained for $x\geq 15$,
with $a=6.13$, $b=15.56$, and $c=12.31$.
By extrapolating the fit,  ${\cal P}(>53.6) = (1.0 \pm 0.6) \times 10^{-6}$ is derived.
On the other hand, the period range used in Figure~\ref{fig:GIS_Rscan}(b),
8.82--8.90 s, comprises 64 independent Fourier wave numbers.
We tested three values of $\dot P$ 
to produce Figure~\ref{fig:GIS_Rscan},
and varied $\Eb$ from 9.0 to 11.0 keV,
with an estimated effective trial number of $\sim 2$.
In addition, 5 values of $E_{\rm L}$ around 3 keV were tested,
to arrive at the final selection of 2.8 keV.
Regarding conservatively these 5 trials all independent, 
the chance  probability to obtain $\zz \geq 53.6$ 
in a 8.82--8.90 s PG is finally estimated as
$64 \times 3 \times 2 \times  5 \times{\cal P}(>53.6) \sim 1.9 \times 10^{-3}$,
or $\Pch \sim 0.2\%$,
when the PPD correction is applied  and $\R$ is allowed to vary at each $P$.
This $\Pch$ applies to the $P_1'$ peak in Figure~\ref{fig:GIS_Rscan}(a).
It is much lower than that  in \S~\ref{subsubsec:step1},
because of  4.5 times more photons utilized here.

\section*{Acknowledgements}
The present work was  supported in part by the JSPS
grant-in-aid (KAKENHI), number 18K03694.


\bigskip
\bigskip
\begin{thebibliography}{}
\bibitem[Aragona et al.2009()]{Aragona09}
Aragona, C., McSwain, M. V., Grundstrom, E. D., et al. 2009, \apj,  698, 514
 \bibitem[Araya \& Harding(1999)]{Harding99} 
 Araya, R. A., \& Harding, A. K. 1999,
\apj,  517, 334
\bibitem[Bosch-Ramon(2021)]{3body_system}
Bosch-Ramon, V. 2021, \aap, 649, C1
 \bibitem[Brazier(1994)]{Z2_94} 
Brazier, K. T. 1994, \mnras, 268, 709
\bibitem[Casares et al.(2011)]{Casares11}
Casares, J., Corral-Santana, J. M., Herrero, A., et al. 2011, High-Energy
Emission from Pulsars and their Systems (Berlin: Springer), 559
\bibitem[Dubus(2013)]{Dubus13}
Dubus, G. 2013, Astron. Ap. Review,  21, 64 (2013) 
\bibitem[Kishishita et. al.(2009)]{Kishishita09}
Kishishita, T.,  Tanaka, T., Uchiyama, Y., \& Takahashi T. 2019,
\apjl, 697,  L1
\bibitem[Makishima(2023)]{IAU363} 
Makishima, K., 2023, Proc. IAU, 363, 267
\bibitem[Makishima et al.(2014)]{Makishima14} 
Makishima, K., Enoto, T., Hiraga, J. S., 
 et al. 2014, Phys. Rev. Lett., 112, 171102
\bibitem[Makishima et al.(2016)]{Makishima16}  
Makishima, K., Enoto, T.,  Murakami, H.,
et al. 2016, PASJ,  68, S12
\bibitem[Makisima et al.(2021a)]{Makishima21a} 
Makishima, K., Enoto, T., Yoneda H., \& Odaka, H. 2021b, 
\mnras,  502, 2266
\bibitem[Makishima et al.(1996)]{GIS2} 
Makishima, K., Tashiro, M., Ebisawa, K., et al. 1996,\pasj, 48, 171
\bibitem[Makisima et al.(2021b)]{Makishima21b} 
Makishima, K., Tamba, T., Aizawa, Y.,  
et al. 2021b, \apj,  923, 63
\bibitem[Miyasaka et al.(2013)]{Miyasaka13} 
Miyasaka, H., Bachetti M., Harrison, F. A., et al. 2013, 
\apj, 775, 65
\bibitem[Ohashi et al.(1996)]{GIS1} 
Ohashi, T., Ebisawa, K., Fukazawa, Y., et al. 1996, \pasj, 48, 157
\bibitem[Serlemitsosn et al.(1995)]{XRT} 
Serlemitsos, P. J.,  Jalota, L.,  Soong, Y., et al.  1995, \pasj,  47, 105
\bibitem[Takahashi et al.(2009)]{Takahashi09} 
Takahashi, T., Kishishita, T., Uchiyama, Y., et al.  2009, \apj,  697, 592
\bibitem[Tanaka et al.(1994)]{Tanaka94}
Tanaka, Y., Inoue H., \& Holt, S. S. 1994, PASJ,  46, L37
\bibitem[Volkov et al.(2021)]{Volkov21}
Volkov, I.,  Kargaltsev., O, Younes, G.,  
et al. 2021, \apj, 615, 61
\bibitem[Yoneda et al.(2023)]{Yoneda23}
Yoneda, H., Bosch-Ramon, V., Enoto, T., 
et al. 2023,
\apj, 948, 77
\bibitem[Yoneda et al.(2021)]{Yoneda21}
Yoneda, H., Khangulyan, D, Enoto, T., et al. 2021, \apj, 917,  id.90
\bibitem[Yoneda et al.(2020)]{Yoneda20}
Yoneda, H., Makishima, K., Enoto, T., 
et al. 2020, Phys. Rev. Lett., 125, id.111103 (Paper I)
\bibitem[Yu et al.(2013)]{glitches}
Yu, M., Manchester, R. N., Hobbs, G., et al. 2013, \mnras, 429, 688
\end{thebibliography}
\end{document}